\documentclass[twocolumn,trackchanges]{aastex631}

\usepackage{amstext,amsmath}

\usepackage[T1]{fontenc}
\usepackage[figure,figure*]{hypcap}
\usepackage{comment}
\usepackage{physics}
\usepackage{soul} 

\usepackage[bb=boondox]{mathalfa}
\usepackage{graphicx}
\usepackage{multirow}
\usepackage{subfig}
\usepackage{colortbl}
\usepackage{float}
\usepackage{enumerate} 

\begin{document}

\title{Results from NASA's First Radio Telescope on the Moon: Terrestrial Technosignatures and the Low-Frequency Galactic Background Observed by ROLSES-1 Onboard the Odysseus Lander}

\correspondingauthor{Joshua J. Hibbard}
\author[0000-0002-9377-5133]{Joshua J. Hibbard}
\affiliation{Center for Astrophysics and Space Astronomy, Department of Astrophysical and Planetary Science, University of Colorado Boulder, CO 80309, USA}
\email{Joshua.Hibbard@colorado.edu}

\author[0000-0002-4468-2117]{Jack~O.~Burns}
\affiliation{Center for Astrophysics and Space Astronomy, Department of Astrophysical and Planetary Science, University of Colorado Boulder, CO 80309, USA}

\author{Robert MacDowall}
\affiliation{NASA Goddard Space Flight Center, 8800 Greenbelt Rd, Greenbelt, MD 20771, USA}

\author{Natchimuthuk Gopalswamy}
\affiliation{NASA Goddard Space Flight Center, 8800 Greenbelt Rd, Greenbelt, MD 20771, USA}

\author{Scott A. Boardsen}
\affiliation{NASA Goddard Space Flight Center, 8800 Greenbelt Rd, Greenbelt, MD 20771, USA}

\author{William Farrell}
\affiliation{DeepSpace Technologies Inc, Columbia, MD 21045, USA}

\author{Damon Bradley}
\affiliation{DeepSpace Technologies Inc, Columbia, MD 21045, USA}

\author{Thomas M. Schulszas}
\affiliation{NASA Goddard Space Flight Center, 8800 Greenbelt Rd, Greenbelt, MD 20771, USA}

\author[0000-0002-3292-9784]{Johnny Dorigo Jones}
\affiliation{Center for Astrophysics and Space Astronomy, Department of Astrophysical and Planetary Science, University of Colorado Boulder, CO 80309, USA}

\author[0000-0003-2196-6675]{David Rapetti}
\affiliation{NASA Ames Research Center, Moffett Field, CA 94035, USA}
\affiliation{Research Institute for Advanced Computer Science, Universities Space Research Association, Washington, DC 20024, USA}
\affiliation{Center for Astrophysics and Space Astronomy, Department of Astrophysical and Planetary Science, University of Colorado Boulder, CO 80309, USA}

\author[0000-0001-7836-1787]{Jake D. Turner}
\affiliation{Department of Astronomy and Carl Sagan Institute, Cornell University, Ithaca, New York 14853, USA}

\begin{abstract}
    Radiowave Observations on the Lunar Surface of the photo-Electron Sheath instrument (ROLSES-1) onboard the Intuitive Machines' \textit{Odysseus} lunar lander represents NASA's first radio telescope on the Moon, and the first United States spacecraft landing on the lunar surface in five decades. Despite a host of challenges, ROLSES-1 managed to collect a small amount of data over fractions of one day during cruise phase and two days on the lunar surface with four monopole stacer antennas that were in a non-ideal deployment. All antennas recorded shortwave radio transmissions breaking through the Earth's ionosphere---or terrestrial technosignatures---from spectral and raw waveform data. These technosignatures appear to be modulated by density fluctuations in the Earth's ionosphere and could be used as markers when searching for extraterrestrial intelligence from habitable exoplanets. After data reduction and marshaling a host of statistical and sampling techniques, five minutes of raw waveforms from the least noisy antenna were used to generate covariances constraining both the antenna parameters and the amplitude of the low-frequency isotropic galactic spectrum. ROLSES-2 and LuSEE-Night, both lunar radio telescopes launching later in the decade, will have significant upgrades from ROLSES-1 and will be set to take unprecedented measurements of the low-frequency sky, lunar surface, and constrain the cosmological 21-cm signal. ROLSES-1 represents a trailblazer for lunar radio telescopes, and many of the statistical tools and data reduction techniques presented in this work will be invaluable for upcoming lunar radio telescope missions.
\end{abstract}

\section{Introduction} \label{sec:intro}
On February 15, 2024, the Intuitive Machines 1 (IM-1) lander named \textit{Odysseus} launched aboard a SpaceX Falcon 9 rocket en route to the Moon, landing approximately a week later on February 22 near the Malapert A crater within 10$^\circ$ of the Moon's south pole. Odysseus, like its namesake, faced and overcame a host of challenges while voyaging to its final destination. One of the main payloads that it carried was NASA's first radio telescope on the Moon, called the \textit{Radiowave Observations on the Lunar Surface of the photo-Electron Sheath}, or \textit{ROLSES-1} (see \cite{burns_low_2021}; Gopalswamy et al. 2025 in prep). This mission was the first of NASA's Commercial Lunar Payload Services \citep[CLPS,][]{burns_low_2021} program to reach the surface of the Moon. 

\begin{figure*}
    \centering
    \includegraphics[width=0.49\textwidth,height=8cm]{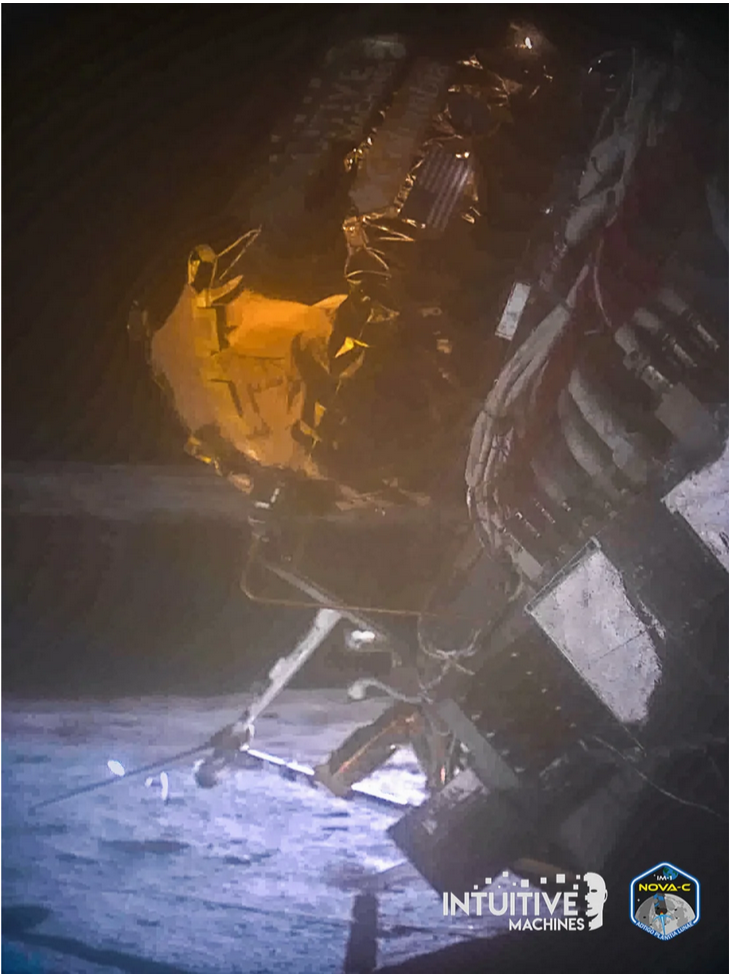}
    \includegraphics[width=0.49\textwidth]{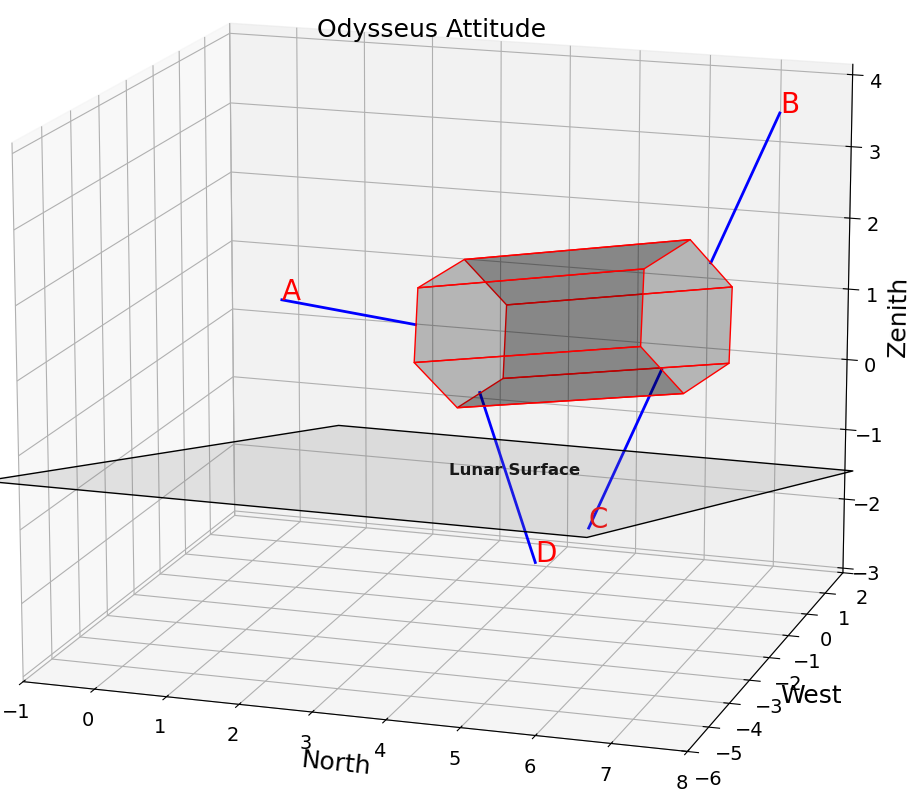}
    \caption{The left panel shows the final image of Odysseus transmitted back to Earth, where the clear tipping of the spacecraft relative to the lunar surface can be seen (image credit to Intuitive Machines). The right panel shows a three-dimensional projection (in arbitrary units) of the Odysseus lander upon the lunar surface, with North and West relative to the Moon's local topography. The ROLSES-1 antennas are labeled from A through D, shown as blue lines with red labels. Odysseus most likely lays upon its North-west panel, crushing antenna D beneath it. The lunar surface slopes to the North at an angle of twelve degrees. The exact height of Odysseus above the lunar surface is not known, and the attitude angles (yaw, pitch, roll) all have five degree errors.}
    \label{fig:odi-orientation}
\end{figure*}

\textit{Odysseus} had a total mass of $1,900~$kg and carried six scientific NASA-payloads along with several commercial ones. It was also the first spacecraft to use liquid methane and liquid oxygen (methalox) propulsion both to the lunar surface and for landing. While in lunar orbit before the planned landing, it was discovered that the laser guidance system was offline, and subsequently using camera systems for navigation \textit{Odysseus} landed with 3 m/s of horizontal velocity and 1 m/s of extra vertical velocity. Upon contact with the lunar surface, the spacecraft tipped over at an angle of $\approx$30$^\circ$  (see left panel of Figure \ref{fig:odi-orientation}), which strongly limited the available spacecraft power and availability of downlink/uplink time.

Despite these circumstances, \textit{Odysseus} maintained power for several more days on the lunar surface and telemetered back data via low gain communication antennas for all scientific instruments onboard, before losing power and shutting down on February 29. \textit{Odysseus} thus not only survived, but endured. ROLSES-1 is a pioneering instrument on the Moon representing decades of effort, research, and technology development undertaken and ventured by hundreds of scientists and engineers to build a lunar observatory \citep{burns_lunar_1990, burns_review_1990, basart_very_1990, johnson_lunar_1990, burns_lunar_1992, johnson_developing_1992, johnson_engineering_1992, burns_low_1994, duric_very_1994, johnson_lunar_1995, lazio_lunar_2009, burns_probing_2012, burns_transformative_2021, karapakula_architecture_2024}.

This first NASA-funded radio telescope on the Moon consisted of four monopole antennas designed to study the low-frequency radio sky and the local lunar plasma wave environment between 5 kHz and 30 MHz. The ROLSES-1 science objectives included: (1) Measure plasma wave activity in the near-surface sheath region; (2) Observe solar and planetary radio waves from a lunar surface observatory, especially now heading into the active solar maximum period; (3) Detect terrestrial natural auroral and human-made radio emissions, to thus assess the Earth as a ``noisy'' radio source; (4) Sense interplanetary and ``slow-moving'' lunar dust via grain contacts with the antenna; (5) Search for evidence of detection of galactic background radiation; (6) Assess the radio frequency interference (RFI) from local sources including the lander as a noise source. In this work, we specifically focus on objectives (3) terrestrial human-made radio noise and (5) the search for the weak galactic background signal, while assessing (6) RFI noise from local sources. 



ROLSES-1 had a primary mode of data collection that transformed raw waveforms into spectra via a digital signal processor (DSP). The DSP data consisted of spectra in two bands, denoted low and high, that ranged from $0.06-1.8$~MHz and from $0.06-30$~MHz respectively, with 512 evenly-spaced channels for each band. The DSP mode provides information about the local radio environment on the lunar surface, including plasma waves at the lunar surface, radio waves from solar bursts, Jovian bursts and emission, short-wave breakthrough transmissions---technosignatures---from the Earth, and the galaxy, the latter of which consists primarily of synchrotron radiation produced by relativistic electrons spiraling in the galactic magnetic field, free-free emission, and free-free absorption by cold electrons in the interstellar medium \citep{zheng_improved_2017,cong_ultralong-wavelength_2021}. DSP data were collected for 83 minutes while in cruise phase and 36 minutes on the lunar surface.

Beside a spectral DSP mode, waveform data can be taken directly from the system's analog-digital converters. The collection of waveforms occurs immediately after the $f_s = 120$~M Samples/second digitization of the measurements. These collections are in short bursts of 4096 samples. However, the mode can in principle be run for tens of minutes allowing the collection of numerous 4096 sample sets which can be integrated to improve resolution. These waveforms can also undergo spectral analysis using ground-based systems for comparisons to the spectra produced on ROLSES-1. On the second lunar day, 27 to 28 waveforms were collected per antenna, with 4096 samples per waveform; no waveform data were collected during cruise phase or the first lunar day. 

Unfortunately, no bursts from either the Sun or Jupiter corresponded with data taken by ROLSES-1 during its observations on three distinct days, according to simultaneous observations made with the STEREO/WAVES spacecraft and from NenuFar in France. Thus, calibration and source comparison between these spacecraft are not possible, and we rely on laboratory measurements and nominal requirements for the antennas and electronics to reduce, clean, and calibrate the ROLSES-1 data. Luckily, a small amount of data were taken by the Owens Valley Observatory's Long Wavelength Array in North America during the cruise phase of ROLSES-1, and we present a comparison of this data to that captured by ROLSES-1 in Appendix \ref{app-lwa}. Another source of comparable data during the cruise phase were taken by a Radio Jove experiment in North America (see Gopalswamy et al. 2025 in prep.).

ROLSES-1 represents a trailblazing effort to place radio telescopes on the lunar surface for NASA, and through the CLPS program several other radio telescopes are scheduled to land on the Moon within the next several years. These telescopes include ROLSES-2, which will be an upgrade to its predecessor with full Stokes parameters, an onboard calibration source, and greater internal electromagnetic interference (EMI) shielding, and the Lunar Surface Electromagnetics Experiment, or LuSEE-Night \citep{bale_lusee_2023}. The latter experiment is designed to land on the farside of the Moon and operate for two years, with a primary science goal of placing the first limits on the cosmological 21-cm line of the Dark Ages of the Universe \citep{pritchard_21-cm_2012}. Both of these radio telescopes will face challenges similar to ROLSES-1, and therefore a large part of this work is to develop tools and statistical techniques that will be useful for reducing, calibrating, and analyzing the data of these upcoming lunar radio telescopes. 

In total, roughly two hours of data were taken by ROLSES-1 and telemetered back to Earth, both spectral DSP and raw waveform data, from all four antennas. In Section \ref{sec:data} we describe these data as well as their cleaning and calibration, and in Section \ref{sec:gal-spec-antenna} we lay out our methodology for fitting the galactic spectrum amplitude and ROLSES-1 antenna parameters using the raw waveform data. Then in Section \ref{sec:results} we discuss the main results from our ROLSES-1 data analyses, including detection of shortwave transmissions from the Earth and the galactic background, and finally we conclude in Section \ref{sec:conclusions}.

\begin{table*}[]
    \centering
    \begin{tabular}{c|c|c|c|c|c|c|c|c|c|c|c}
        \hline
         \multicolumn{3}{c}{} & \multicolumn{3}{|c|}{Status} & \multicolumn{3}{c|}{Remaining Channels (HB)} & \multicolumn{3}{c}{Remaining Channels (LB)} \\
         \hline
         Antenna & Panel & Gain & In-Transit & LS1 & LS2 & In-Transit & LS1 & LS2 & In-Transit & LS1 & LS2 \\ & & & $80$~min & $10$~min & $25$~min & & & & & & \\ \hline \hline 
         A & North-East & Low & St & St & De & 0 & 0 & 6 & 0 & 7 & 0 \\ B & South & High & St & De & De & 0 & 5 & 23 & 0 & 3 & 2 \\ C & North & High & St & St & De & 0 & 0 & 3 & 0 & 0 & 0 \\ D & North-West & Low & De & Cr & Cr & 64 & 0 & 0 & 4 & 0 & 0 \\ \hline
    \end{tabular}
    \caption{Antenna position and status for each observation. \textit{Panel} denotes upon which Odysseus panel the antenna resides, while \textit{Gain} denotes the pre-amp daughter card gain setting. Under \textit{Status}, \textit{St} refers to stowed antennas, \textit{De} to deployed, and \textit{Cr} to an antenna likely crushed by the landing. \textit{Remaining Channels} denotes the number of channels with data remaining out of the 512 (histogram) frequency channels after the six-step cleaning process for both the high and low band (HB, LB respectively).}
    \label{tab:data}
\end{table*}

\begin{table}[]
    \centering
    \begin{tabular}{c|c}
        \hline
        Data Level & Reduction \\
        \hline \hline
        1 & Uncalibrated, Raw Data \\ 
        2 & Remove DSP Quantized Noise \\ 
        3 & Remove Contaminated Ground Channels \\
        4 & Remove Contaminated Stowed Channels \\
        5 & Remove DSP Noise and Correct Zero-Level \\
        6 & Calibrate to Voltage \\
        7 & Convert to Power Spectral Density \\
        \hline
    \end{tabular}
    \caption{Table showing the seven data levels and the reduction process implemented at each step.}
    \label{tab:datalevels}
\end{table}

\section{Data} \label{sec:data}

\begin{table}[]
    \centering
    \begin{tabular}{c|c|c}
        \hline
        Label & Date & Time [UTC] \\
        \hline
        \hline
        In-Transit & Feb 21, 2024 & 3:17:46 - 4:37:46 \\
        Lunar Surface 1 (LS1) & Feb 26, 2024 & 17:13:39 - 17:21:39 \\
        Lunar Surface 2 (LS2) & Feb 27, 2024 & 5:39:56 - 5:55:56 \\
        \hline
    \end{tabular}
    \caption{Table showing the label, date, and UTC for each of the three observation days of ROLSES-1.}
    \label{tab:obs-dates}
\end{table}

The ROLSES instrument consists of four telescoping, stacer monopole antennas each $L=2.5$ meters long, with two antennas situated closer to the base of the lander (denoted the lower antennas) and two closer to the top (the upper antennas). If we denote each antenna by its respective position on the \textit{Odysseus} lander, whose North direction nominally faces the Earth (in local lunar coordinates), then there are two upper antennas oriented North and South and two lower antennas oriented North-East and North-West. Table \ref{tab:data} summarizes the locations on the lander, deployment status, and pre-amp card gain setting, and number of channels with data from the digital signal processor (DSP) remaining after data reduction for each of the three days of observation. 

Moreover, Figure \ref{fig:odi-orientation} shows a three-dimensional approximate projection of Odysseus' attitude on the lunar surface, with the gray plane denoting the $12^{\circ}$ slope of the crater Odysseus landed in on the Moon. These attitude angles have five degree uncertainties, and were calculated according to measurements from the gravitometric readings of the fuel tanks. The four antennas are colored in \textcolor{blue}{blue} and labeled with their respective antenna designations. We note that the antenna labeled $D$ is on the same panel that the Odysseus lander rests upon and is likely crushed. Furthermore, thermocouple temperature readings from the housekeeping data for antenna $D$ showed values which were consistent with a broken or shorted antenna. Although the exact height of Odysseus above the lunar surface depends upon the helium tank and state of the lander legs--which are not known exactly--it is possible that Antenna $C$ is touching the lunar regolith (as shown in Figure \ref{fig:odi-orientation}), although its temperature readings were consistent with a functioning antenna. 

There are three distinct days during which data were collected from the ROLSES-1 antennas aboard Odysseus; the label, date, and UTC for each observation day are shown in Table \ref{tab:obs-dates}. While en route to the Moon, and in order to offset a helium tank leak on the lander, the roll maneuvers which change the panel facing the sun were ceased. As a result, antenna $D$ became sun-facing and heated to at least $80^{\circ}$~C. It was then ejected from its housing. This however turned out to be a fortuitous occurrence, as it allowed antenna $D$ to collect nearly eighty minutes of data while in cis-lunar space. This observing day was on $02/21/24$ and is denoted as \textit{In-Transit} in all subsequent data analyses. After landing, Antenna $B$ also reached the critical temperature and ejected from its housing, allowing engineers to collect data with it for approximately ten minutes on $02/26/24$, denoted in this analysis as \textit{LS1} for \textit{Lunar Surface day 1}. 

\begin{figure*}
    \centering
    \includegraphics[width=0.48\textwidth]{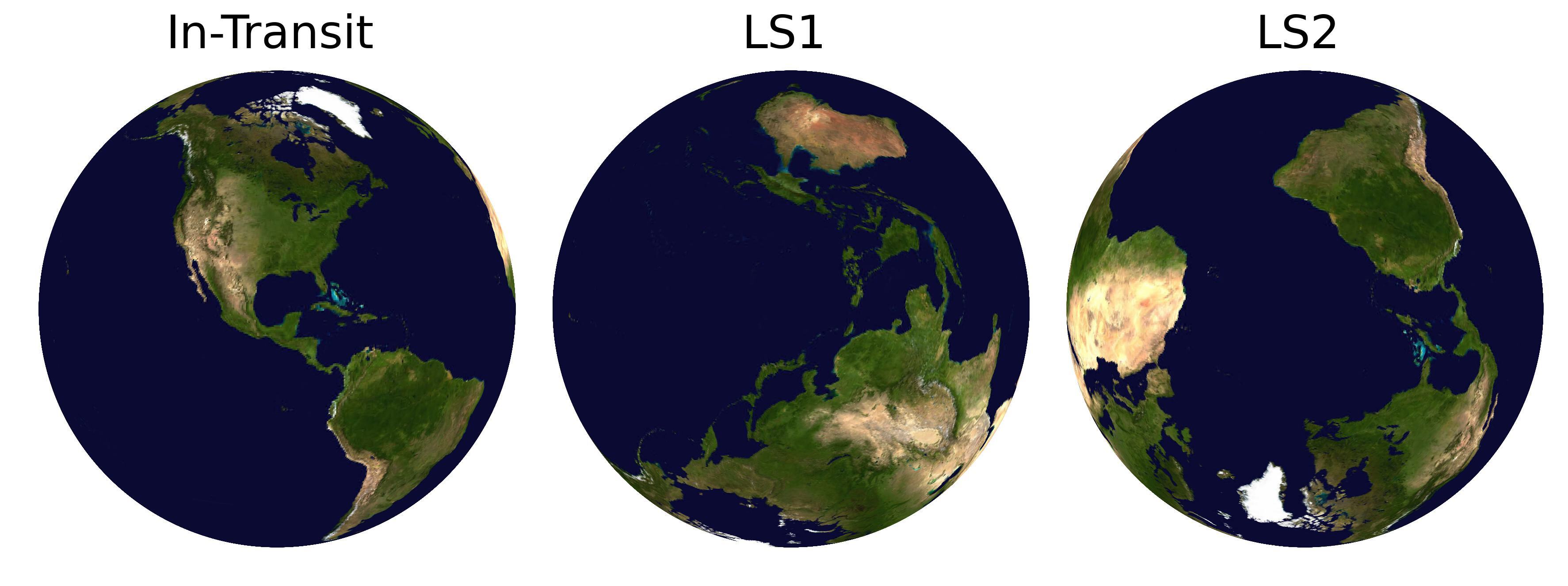}
    \caption{Approximate view of the Earth as seen by Odysseus during each of the three observing days: In-Transit (02/21/24), Lunar Surface 1 (02/26/24), and Lunar Surface 2 (02/27/24). Note that the entire North American continent is in-view during the In-Transit observations, while two large bodies of water (the Pacific and Atlantic ocean) are largely in-view during the subsequent two observation days.}
    \label{fig:earth-orientation-obs}
\end{figure*}

On $02/27/24$, the command to the ROLSES-1 computer module was given for the final two antennas $A$ and $C$ to deploy, which they did successfully and subsequently recorded twenty-five minutes of data from the three functioning antennas. We denote this day \textit{LS2}, and note that five minutes of raw waveform data were also collected on this day with all three functioning antennas. Figure \ref{fig:earth-orientation-obs} shows an approximate view of Earth during each of the three days of observation. Note that North America was in full view during the In-Transit data collection, while LS1 and LS2 views of Earth are dominated by oceans (the Pacific and Atlantic, respectively). According to THEMIS-ARTEMIS data\footnote{\url{ https://artemis.igpp.ucla.edu/overview_data.shtml}}, the spacecraft was in the magnetosheath for all three observing days. 

To reiterate, ROLSES-1 had two primary modes of data collection: one which telemeters back raw waveform packets in the time-domain (the waveform mode) for all antennas every $16$~s; and a second mode which telemeters back spectra produced by the DSP (spectral mode) for all antennas every $8$~s. The DSP measures data in both a high-band and low-band. The high band frequencies range from $0.06 - 30$~MHz with 512 evenly-spaced channels and a channel width of $\Delta \nu_H = 58.594$~kHz. To measure in the low-band the DSP down-samples by a factor of sixteen, resulting in channels from $0.06 - 1.87$~MHz with 512 evenly-spaced channels and a channel width of $\Delta \nu_L = 3.662$~kHz. For this work, which focuses on the data analysis of ROLSES-1, we only give a brief review of the relevant aspects of the DSP. More details on the specifications of ROLSES-1, its parts, and its engineering, can be found in Gopalswamy et al. (in prep). 

Regarding onboard spectral processing, a low-pass filter is applied to the voltage time series measurements, gradually suppressing frequencies above the science band maximum of $30$~MHz. The filtered waveforms are then tapered by a windowing function after which the DSP averages the absolute value squared of the discrete Fourier transform (DFT) of each windowed waveform. The window of the DFT overlaps by one half with each subsequent DFT of the next waveform. A bit slicer was utilized within the DSP in order to reduce the number of bits telemetered back to Earth, as not all of the 32-bit range would be needed to encode the DFT data. The bit slicer therefore sampled 16 bits from the full 32-bit range, effectively encoding a dynamic range for the data: incoming waves (of a particular frequency) below the minimum threshold of this dynamic range would not trigger the DSP to register a measurement, and only quantized noise values would be output. Channels which do fire appear within the dynamic range of the bit slicer and then respond linearly to input power until reaching a maximum threshold, above which the DSP output saturates. Upon examination of the raw waveforms taken during ground testing (see Section~\ref{sec:groundtest}) and on LS2, which have count values of several hundreds, well below the $14$-bit ADC range, it does not appear that this saturation level was ever reached. The lower threshold, however, was often not reached, and we see evidence of quantized noise in the DSP data (see below).


\begin{figure*}
    \centering
    \includegraphics[width=0.48\textwidth]{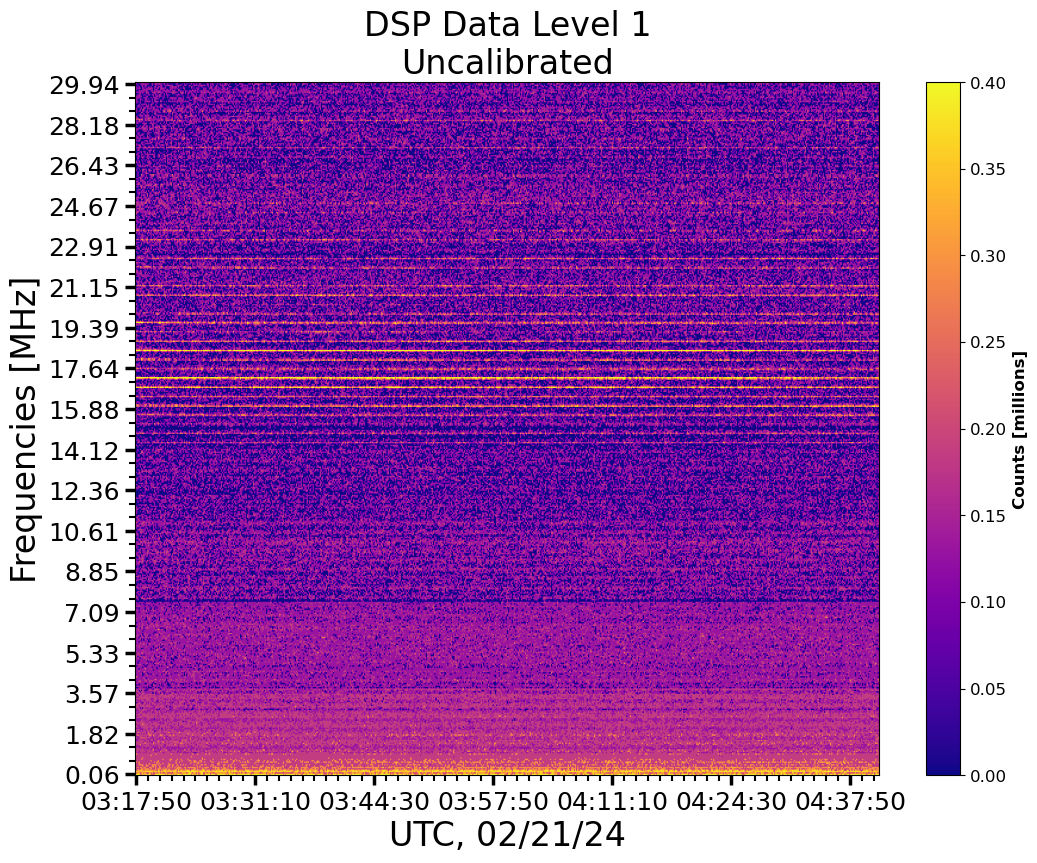}
    \includegraphics[width=0.48\textwidth]{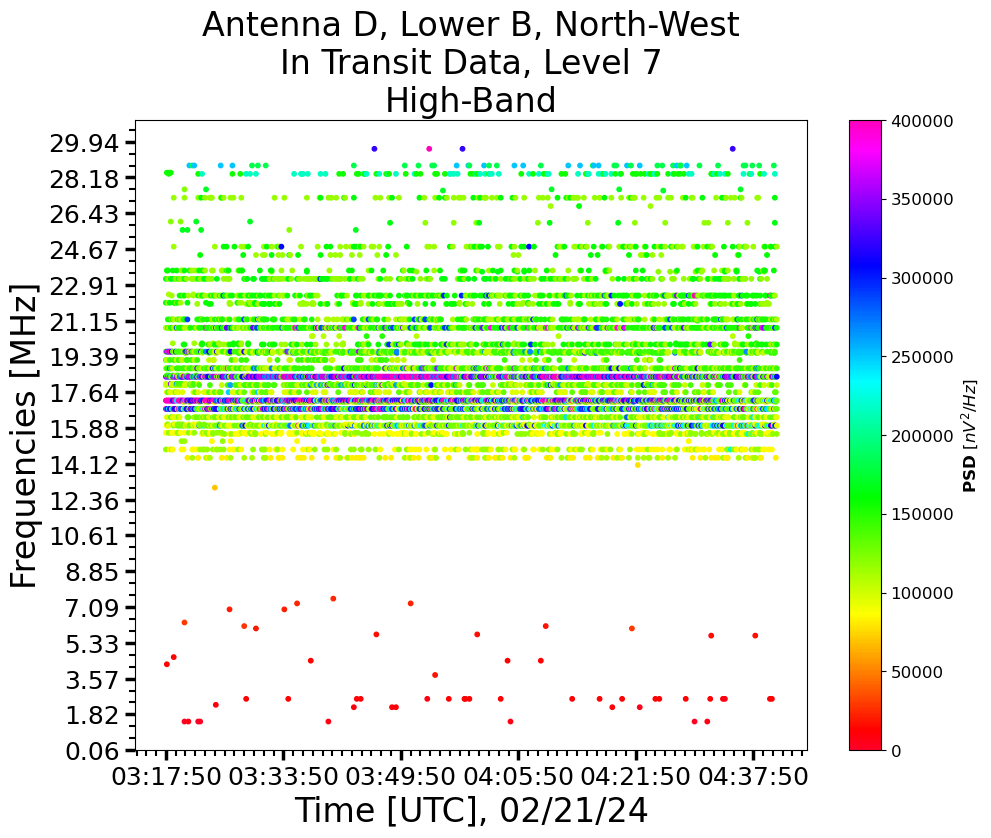}
    \caption{Dynamic spectra or waterfall plots for DSP data. The left panel shows the uncalibrated, un-cleaned level 1 data, while the right panel shows the result of the entire cleaning and calibration process, resulting in data of level 7. These DSP data correspond to the high band, from $0.06-30$~MHz, and for antenna D during the In-Transit observation day. Units for the left panel are raw counts, while the right panel is shown in $n V^2/Hz$. All times are in UTC.}
    \label{fig:dsp-waterfall-intransit}
\end{figure*}

\begin{figure*}
    \centering
    \includegraphics[width=0.48\textwidth,height=7
    cm]{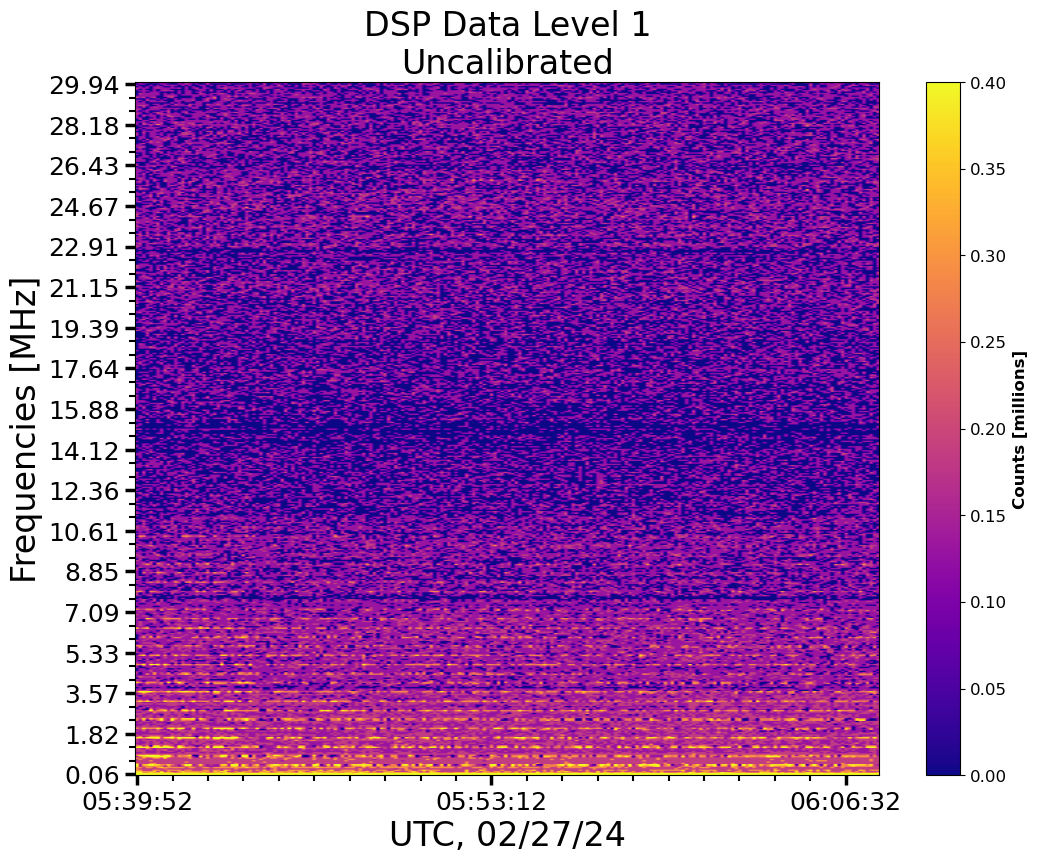}
    \includegraphics[width=0.48\textwidth,height=7cm]{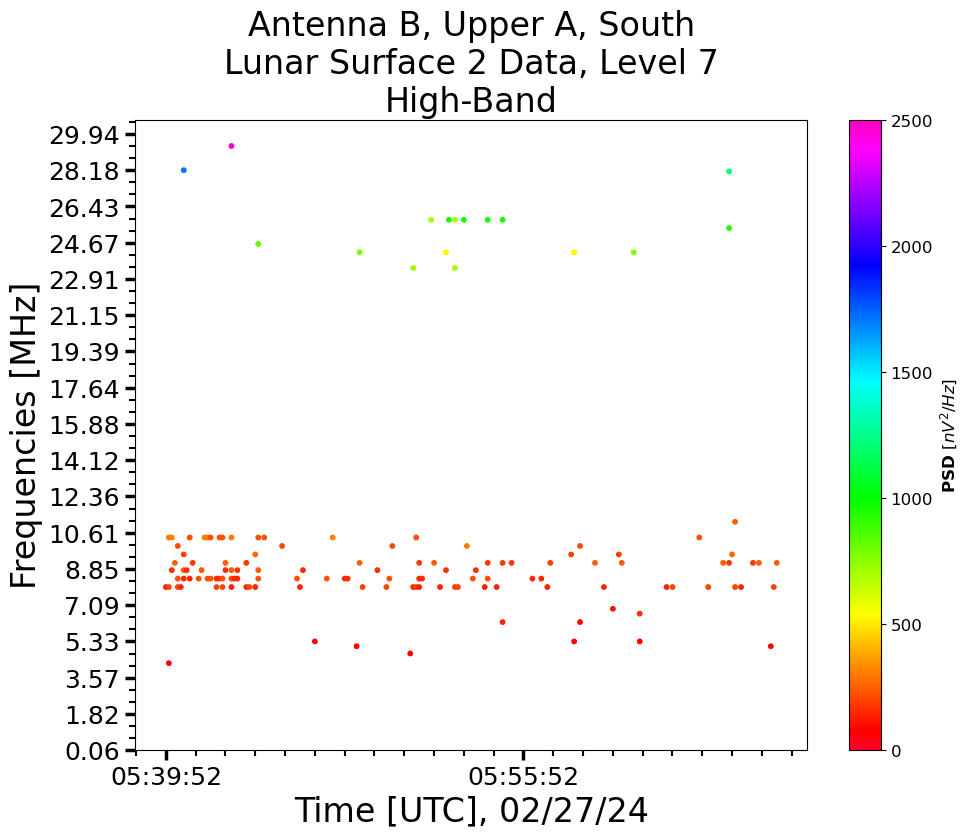}
    \caption{Waterfall plots for DSP data. Same as Figure \ref{fig:dsp-waterfall-intransit}, but for the LS2 observation day and antenna B.}
    \label{fig:dsp-waterfall-ls2}
\end{figure*}

The raw waveform data consist of noise from the front-end electronics (FEE) and lander, the galaxy, and any radio frequency interference (RFI) from Earth in the form of shortwave transmissions breaking through the Earth's ionosphere (the terrestrial technosignatures to be discussed more extensively in Section \ref{sec:results}). These three noises can be written as
\begin{equation}
    \hat{\sigma}_\mathrm{wf}(\nu) = \hat{\sigma}_\mathrm{FEE}(\nu) + \hat{\sigma}_\mathrm{g}(\nu) + \hat{\sigma}_\mathrm{st}(\nu),
\end{equation}
where the \textit{hat} symbol reminds us that these are random variables and not deterministic components, and all are functions of the observing frequency $\nu$. In contrast, the noise in the DSP output is either quantized noise for channels which do not trigger, represented as $\hat{\sigma}_\mathrm{quant}$, or can be written as
\begin{equation}
    \hat{\sigma}_\mathrm{DSP}(\nu) = \hat{\sigma}_\mathrm{wf}(\nu) + \hat{\sigma}_\mathrm{lin}(\nu),
\end{equation}
for channels which are triggered. The term $\hat{\sigma}_\mathrm{lin}$ represents the noise added by the DSP for a channel in the linear response regime when the DSP is triggered.

\subsection{Ground Test Data}
\label{sec:groundtest}

Lastly, in December of 2023 after integration of ROLSES-1 onto Odysseus but before final integration of the lander onto the rocket, the lander and ROLSES-1 were switched on. Data were collected for all four stowed antennas and both data modes, spectral and waveform. These test data are denoted as \textit{Ground Data} and are used in the upcoming analyses for comparison against the data taken in cis-lunar space and on the lunar surface.

\subsection{Data Collection Modes}

Each data collection mode exhibits different sources of noise, error, and challenges to data reduction. We discuss now the data reduction process for each collection mode.

\subsubsection{DSP Spectral Mode Data}

\begin{figure}
    \centering
    \includegraphics[width=0.47\textwidth]{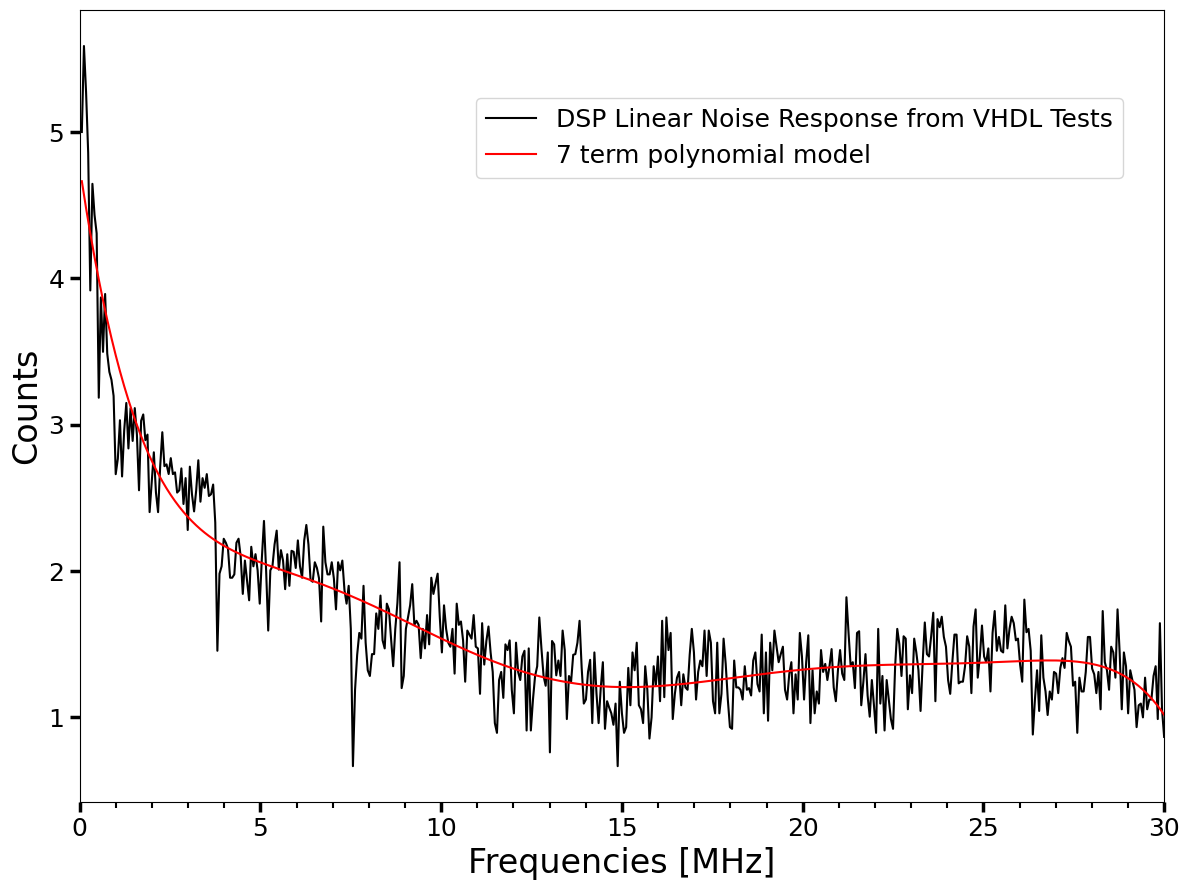}
    \caption{Plot showing the DSP linear noise response as generated from a test-bench simulation (using VHDL or VHSIC Hardware Description Language) for ROLSES-1 at NASA GSFC. The black curve shows the mean of thirty input signals near the lowest detectable level of the DSP, scaled according to the DSP output response. The red curve shows a fit from a seven-term polynomial to the black data.}
    \label{fig:dsp-lin-noise}
\end{figure}

We define Level 1 to be the uncalibrated, uncleaned data telemetered back to Earth from the ROLSES-1 DSP. Waterfall plots for the In-Transit and LS2 uncalibrated data are shown in the left panels of Figures \ref{fig:dsp-waterfall-intransit} and \ref{fig:dsp-waterfall-ls2}, respectively, with frequencies in MHz and time shown in Coordinated Universal Time (UTC) for each day. Unfortunately the cleaning process removed nearly all data from LS1, so it is not shown here. 

There are six ensuing steps in the data cleaning and calibration process, summarized in Table \ref{tab:datalevels}. In detail, each data level consists of the following procedures:

\begin{enumerate}
    \item \textbf{Level 1:} Uncalibrated, uncleaned DSP data.
    \item \textbf{Level 2:} Remove all quantized noise, $\hat{\sigma}_\mathrm{quant}$. This involves setting any channel to zero if its value is equal to or below the maximum quantized noise value. Note that this step alone eliminates most of the Ground test data.
    \item \textbf{Level 3:} Flag remaining channels in the Ground test data, and remove these flagged channels from all the data. These are channels considered contaminated with FEE noise, $\hat{\sigma}_\mathrm{FEE}$, or the lander itself. Therefore, this step removes the remainder of the channels in the Ground test data.
    \item \textbf{Level 4:} Flag and remove channels in stowed antennas for data collected after launch but before deployment. If an antenna is stowed, and one of its channels triggers, we consider that a contaminated channel and remove it from all the data. No data remains now for antennas which were stowed during data collection.
    \item \textbf{Level 5:} Remove DSP noise in the linear response regime, $\hat{\sigma}_\mathrm{lin}(\nu)$. This noise level is estimated using simulations from a ROLSES-1 test-bench model at NASA Goddard Space Flight Center (GSFC). A broadband white-noise signal is input to the model that is just enough to cause the channels in the DSP to trigger. This simulation is run at least thirty times to suitably characterize the distribution, and the average of them is shown with the black line in Figure \ref{fig:dsp-lin-noise}. We then fit a seven term polynomial (shown by the red curve) to the simulations and approximate this curve as the linear response noise of the DSP. This curve is then subtracted from the remaining data.
    \item \textbf{Level 6:} Convert DSP counts to voltages using calibration sweep lab data. This test consisted of injecting signals at known voltage levels at frequencies across the band and measuring the resultant counts. We then use bi-linear interpolation to fill in values between frequency channels and between voltage measurements from the lab calibration tests, giving us calibrated voltages $V_\mathrm{DSP}[f]$. It should be noted, however, that this calibration does not take into account the effects of losses due to the antennas, their coupling to the local environment, or their directivity, but applies directly to the response of the front-end electronics and pre-amp/daughter cards.
    \item \textbf{Level 7:} Convert calibrated volts to power spectral density (PSD, with units of $V^2/Hz$) via 
    \begin{equation}
        S_\mathrm{DSP}[f] = \frac{2}{f_s} V^2_\mathrm{DSP}[f],
    \end{equation}
    where the factor of two arises from the one-sided Fourier transform convention, and we divide by the sampling frequency $f_s$ instead of the bandwidth $\Delta f$ due to the Fourier transform normalization convention used in Simulink (see Appendix \ref{app-psd}).
\end{enumerate}

In theory, the noise remaining in the DSP now consists approximately of remote sensed signals like the galactic noise and shortwave transmissions. However, residual electronic noise due to changes in the response of the FEE from the time of the Ground tests to those of the observations, along with any other noise produced by the coupling of the system to the lunar surface, as well as noise from the spacecraft produced by both radiative and conductive coupling, are probably present since removal of these noise sources was not addressed in the analysis. These sources of noise are difficult to assess, and are probably temperature dependent. Furthermore, there will be a small capacitive coupling of the antenna to the photoelectron sheath of the Moon, and to the sheath formed at the antenna itself, separate from the former. These couplings however are expected to be small \citep{farrell_complex_2007,farrell_lunar_2013} on the South Pole due to the low angle of the Sun on the lunar horizon and are not an issue above the local electron plasma frequency.

The right panels of Figures \ref{fig:dsp-waterfall-intransit} and \ref{fig:dsp-waterfall-ls2} show the ``dot'' waterfall plots for the fully cleaned and calibrated data at Level 7 in units of PSD. Note that Figure \ref{fig:dsp-waterfall-intransit} is for Antenna $D$, deployed in transit, and Figure \ref{fig:dsp-waterfall-ls2} is for Antenna B, which is oriented nearly perpendicular to the lunar surface (see Figure \ref{fig:earth-orientation-obs}).

The brightest lines for the In-Transit data shown on the right panel of Figure \ref{fig:dsp-waterfall-intransit} are those colored in magenta and blue around $16.4, 18.2,$ and $19.4$~MHz with PSD values around $\sim 3.5 \times 10^5  nV^2/Hz$, and they also appear to be the most stable over the 80 minutes of data collection. Fainter signals still appear above $27.1$~MHz that are less than $2 \times 10^5 nV^2/Hz$ in PSD, and below $\sim 8$~MHz the signals are around $5 \times 10^4~nV^2/Hz$, although they appear more sporadically, especially near the low-end of the band.

There is also a curious dearth of data between about $8$ and $14$~MHz for the In-Transit observation. However, we do not see any large noise fluctuations in these corresponding frequency ranges of the Ground data for Antenna D, indicating that these channels were not removed due to large noise fluctuations during the cleaning process. Only levels consistent with DSP quantized noise were present, so it would appear that, at least for this observation, there were no signals detected by the antenna in that frequency range which were great enough to trigger the DSP. Whether that is due to the strength of the signals themselves at that UTC and location on Earth (see Figure \ref{fig:earth-orientation-obs}), the conditions of the ionosphere, or an effect of the antenna itself, is difficult to determine\footnote{Although the DSP Ground data for antenna D show only quantized noise, the waveform Ground data for antenna D also exhibit many other noise features, including a similar absence of noise features around the $8-14$~MHz frequency range, as that in the DSP In-Transit data. This could indicate that other noise features on their own were not great enough to trigger the DSP (causing it to only output quantized noise), but when signals such as human-made radio frequency interference from Earth were also present, the combination of noise features and signals was powerful enough to trigger DSP channels. Unfortunately no data exists for antenna D on the lunar surface, so this cannot be verified by comparing waveform data on the lunar surface with the Ground waveform data.}.

Fortunately there exists a snapshot of data taken by the Owens Valley Long Wavelength Array (OVRO-LWA) in North America at a similar UTC for the In-Transit observation day, February 21, 2024. These data from the LWA consist of calibrated autocorrelations from $\sim 0.01 - 90~$MHz for each antenna in the 352 array for a single time stamp at $3:10:20~$UTC, right near the beginning of the ROLSES-1 observations on the corresponding day. Below $30~$MHz we see shortwave transmission signals like those detected by ROLSES-1, with similar bright features in the $14 - 30~$MHz band.  We note also that there is a similar drop-off in signal strengths below $14~$MHz detected by the LWA data, with similar intermittent signals between $3-8~$MHz, further strengthening the conjecture that the drop-off in signals below $14~$MHz is real, as the data between the LWA observatory and ROLSES-1 are consistent. Appendix \ref{app-lwa} presents the LWA data as well as an analysis to estimate the power of the shortwave transmitters as seen by ROLSES-1. Assuming the transmitters have isotropic directivities and that ROLSES-1 can be approximated as a dipole in free space, we find the estimated power of the transmitters to be around several mega Watts for the brightest lines (the blue and violet dots in Figure \ref{fig:dsp-waterfall-intransit}), with the rest in the kilo Watt range (the green dots in Figure \ref{fig:dsp-waterfall-intransit}). These values are broadly consistent with what is expected for shortwave transmitters on Earth, although the most powerful never exceed $\sim 2~$MW. It is likely that our calibration method is therefore off by a small factor: a likely culprit is the zero-level from the calibration sweep lab data. The zero-level is set by the maximum value of quantized noise \textit{observed}, and so there may be some error in where the true zero-level lies. Because the DSP does not trigger above this threshold, this estimate for the zero-level is the best that can be obtained without better calibration and testing data, which is not available for ROLSES-1. These issues are already being understood and corrected for ROLSES-2, the re-fly mission. In addition, if any solar or Jovian bursts had been observed, these data could have been used in conjunction with other spacecraft or ground-based observatories to better help calibrate the ROLSES-1 DSP data.\footnote{Additionally, if the raw waveform data (examined in the next section) could be calibrated absolutely, and gave a small spread of values for the ground test data and thus allowing the zero-level to be well-measured, then comparisons between the waveform and the DSP data could be used for calibration. Unfortunately, no such absolute calibration is possible, and the spread in the ground test data for the raw waveform is broad for antenna D.}

Finally, the right panel of Figure \ref{fig:dsp-waterfall-ls2} shows the dot waterfall for antenna B on observation day LS2. Most of the DSP data has been removed, though some lines still exist in the $\sim 3.57 - 10.61$~MHz range with faint PSD values significantly below $1000 ~nV^2/Hz$. Note also that data were collected for only 25 minutes on LS2. No signals with PSD values greater than $\sim 2500 ~n V^2/Hz$ were detected by antennas B or C during this observation day by the DSP. All DSP data taken by the three surviving antennas on LS2 are summarily shown as cyan ``x's'' in Figure \ref{fig:wf-average}. Note that these data points, which have been just over-plotted, have not been time-averaged (due to the small number of data points), whereas the waveform data (in black and red) have been time-averaged. The good agreement between the cyan DSP data and the black and red waveform data appears to indicate that the antennas and front-end electronics behaved near the nominal gain settings.

\subsubsection{Waveform Mode Data}

\begin{figure*}
    \centering
    \includegraphics[width=0.95\textwidth]{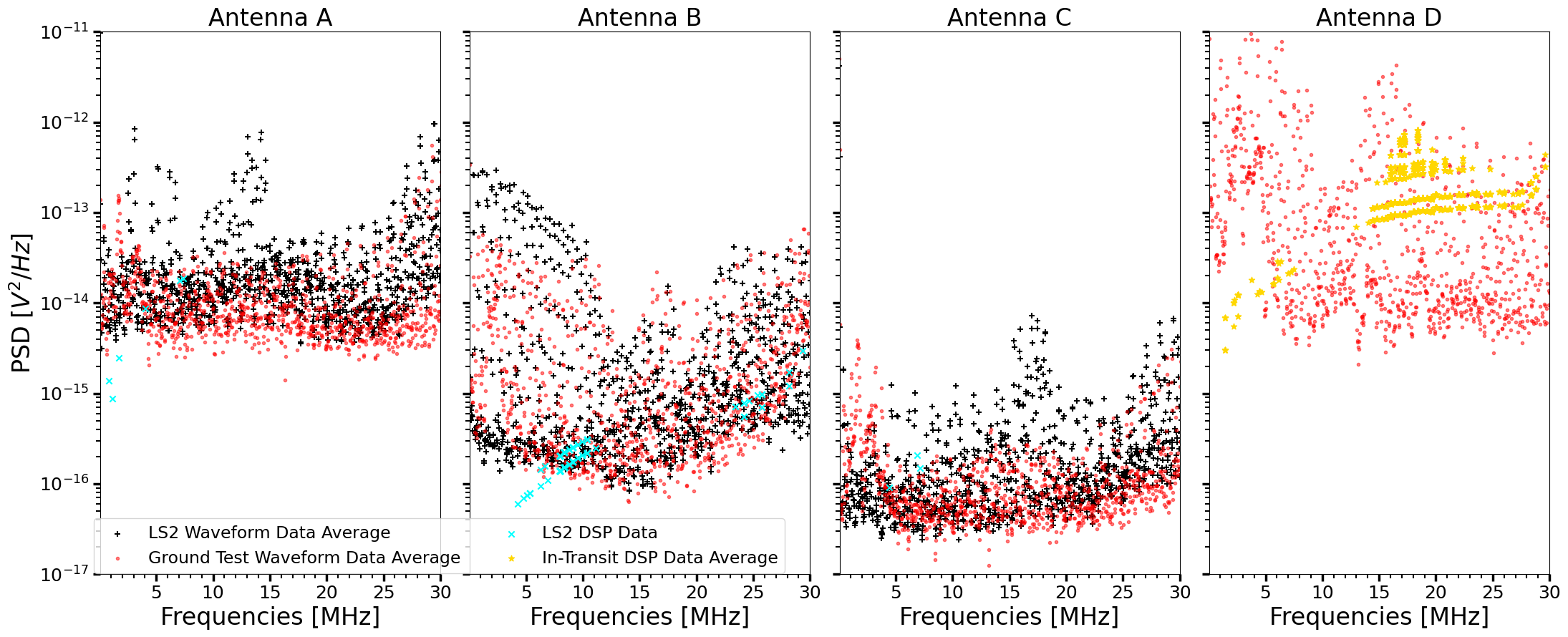}
    \caption{Raw waveform time-averaged spectra for the three antennas that were functional on the lunar surface (A, B, C) after transformation to the time-domain using spectral analysis. Antenna D is included for reference in terms of In-Transit (gold stars) and Ground testing data. The black data points show the time-averaged LS2 data up to $30$~MHz, while the red data points show the equivalent time-averaged spectra from the Ground tests. The cyan crosses show the DSP data from the same observation day for each functional antenna, and these have not been averaged in time due to the paucity of data. The agreement between the nominally calibrated raw waveforms and the laboratory calibrated DSP data is reassuring, and deviations are discussed in subsequent sections.}
    \label{fig:wf-average}
\end{figure*}

The waveform packets each consist of $N=4096$ points in the time-domain taken every 16 seconds for each antenna. As no calibration data exists for these raw waveforms, we rely upon the nominal requirements set for the antennas and pre-amp daughter card gain settings during construction of ROLSES-1. The upper antennas (B and C) were configured in the high gain mode, with a nominal pre-amp daughter card gain value of 36. To avoid potential saturation by the photoelectron sheath's plasma waves, the lower antennas (A and D) were configured in the low gain mode and set to 3.6. Even allowing for appreciable error above and below these nominal gain values, we do not find substantial changes to the resultant PSD levels of the waveform spectra.

The requirements were set for the upper antennas so that the ADC would saturate for an input monochromatic signal or electric field with a value of $\pm 33$~mV/m. The discrete levels in this range, given a 14-bit ADC, then translate to approximately 1 count per electric field with strength $\pm 4 \mu V$/m. The lower antennas, based on the reasoning above, were set to saturate for a signal ten times higher, giving 1 count per electric field with strength $\pm 40 \mu V$/m. These configurations along with the gain settings allow us to convert the counts in the raw waveform packets into nominal voltages, $x_p(n)$.

We then spectrally analyze the 4096 sample waveform set, and apply a windowing function $w_p(n)$ given by a 4th order Blackman-Harris window multiplied by a sinc function to each packet $x_p(n)$ in the time domain to suppress power in the sidelobes and account for the finite length of the waveform. Typically one would then implement an M-component decimating polyphase filterbank (PFB) to each of these tapered packets, where each branch $y_p(n)$ indexed by $p$ is given by
\begin{equation}
    y_p(n) = \sum_{m=0}^{M} w_p(m) x_p(n -m),
\end{equation}
and then form the sum over all branches $P$ after calculating their DFT via: 
\begin{equation}
    \tilde{Y}(k,n) = \sum_{p=0}^{P} \sum_{m=0}^{M} \big[ w_p(m) e^{-2 \pi i k p/P} \big] x_p(n -m).
\end{equation}
However, as we are limited to packets with $N=4096$ points taken 16 seconds apart, we do not have a continuous stream of data of length $N_d \gg N$ to split into several taps ($\sim 4-8$) while also maintaining high frequency resolution. We can though form instead the usual Fourier transform spectrum of our weighted time-stream as
\begin{equation}
    \tilde{X}[f] = \frac{1}{N} \sum^N_{n=0} x[n]w[n]e^{-2 \pi i n f /N}.
\end{equation}
We then calculate power spectral density (PSD) values following \cite{paschmann_analysis_1998} for the Fourier transformed waveform data $\tilde{X}[f]$ by forming the weighted, autocorrelation via
\begin{equation}
    S_w[f] = \frac{2}{\Delta f}\frac{1}{W_\mathrm{ss}} |\tilde{X}[f]|^2
    \label{eq:psd}
\end{equation}
where the factor of two arises from our one-sided Fourier transform convention (i.e., ignoring negative frequencies), $\Delta f = f_s / N$ is our frequency resolution, $f_s=120$~MHz is our sampling frequency, and $W_\mathrm{ss}$ normalizes the PSD to account for the energy loss effects due to the windowing function as follows
\begin{equation}
    W_\mathrm{ss} = \frac{1}{N} \sum^N_{n=0} w^2[n].
\end{equation}
Note that different programs and algorithms use different Fourier transform normalization conventions, and that depending on the convention, the equation to calculate the PSD $S_w$ will be different than that presented in Equation~\ref{eq:psd}. See Appendix \ref{app-psd} for a discussion of the different normalizations and their effect on calculating the PSD.

Figure \ref{fig:wf-average} shows the resultant spectral PSD for each antenna. The black markers show the time average of all spectra for the LS2 waveform data (approximately 5 minutes of data), while the red markers show the equivalent spectra for the Ground waveform data. We have also included all DSP data taken on observation day LS2 as cyan crosses, for which no time-averaging has been implemented. Finally, the gold stars show the time-averaged In-Transit data for Antenna D. We note that Antenna A and C both exhibit clear deviations in the black LS2 data points away from the red Ground data points, particularly in the $\sim 3 - 20$~MHz range. These deviations are discussed along with the DSP data in subsequent sections.

\subsubsection{Data Reduction and Noise Estimate for Galactic Spectrum Studies}
\begin{figure}
    \centering
    \includegraphics[width=0.5\textwidth]{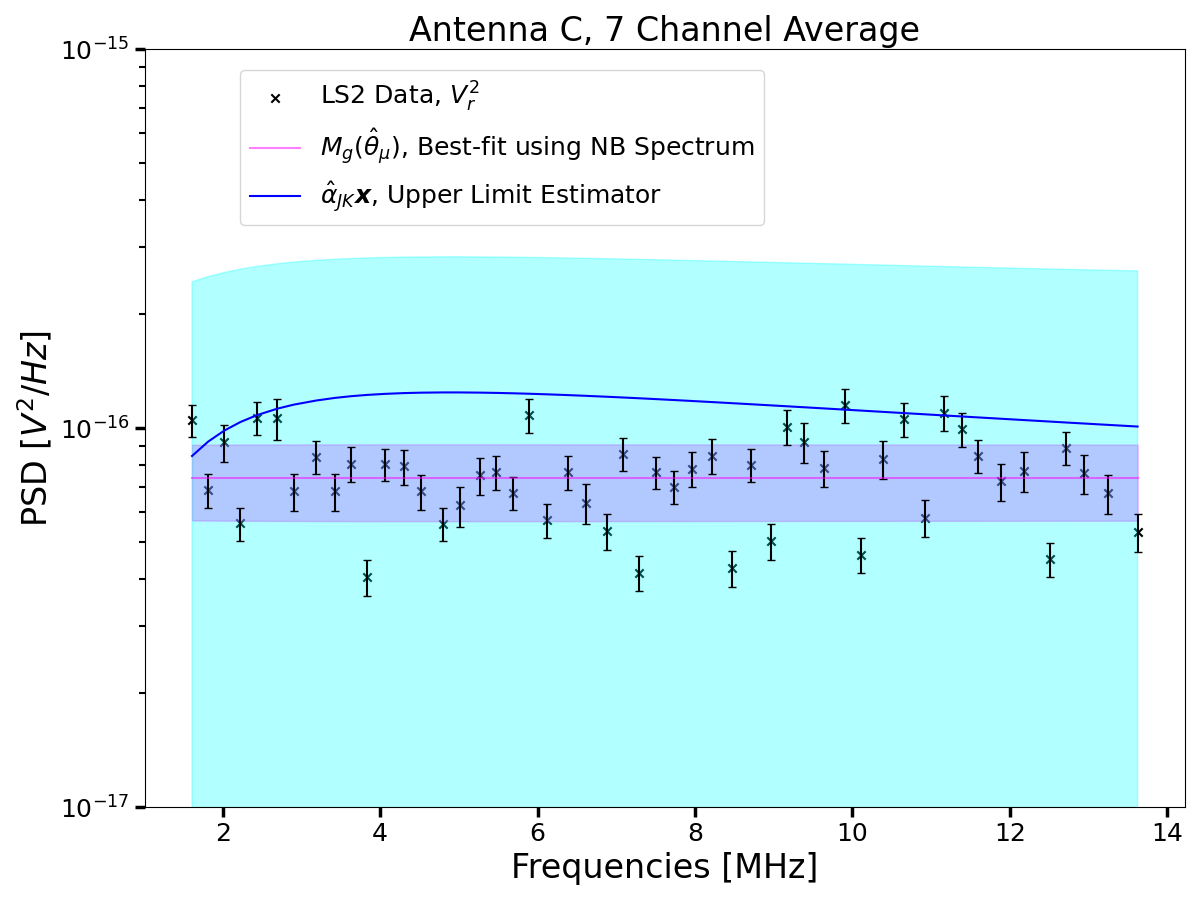}
\caption{PSD data from Antenna C after sigma-clipping and averaging are shown as black crosses with associated error bars; note that after averaging seven adjacent frequency channels together, the noise level in the data is $\sim 100~nV^2/Hz$, although the intrinsic scatter in the data is larger at about $\sim 800~nV^2/Hz$. The dark blue line with a cyan error band corresponds to the galactic spectrum of \cite{novaco_nonthermal_1978} denoted as the vector $\boldsymbol{x}$ multiplied by the jack-knife estimator of the antenna parameters, $\hat{\alpha}$, over the frequency range depicted in this plot. The latter should thus be taken as an upper limit, and the cyan error bands are calculated according to the standard variance in a jack-knife estimate.  The magenta line with a $1\sigma$ purple error band shows a nested sampling fit to the data using the full model $M_g$ that includes noise parameters, antenna parameters, and an amplitude for the galactic spectrum, all represented as the mean parameter vector $\hat{\theta}_{\mu}$. The magenta error band is calculated from the propagating the parameter covariances from the nested sampling fit into PSD data space.}
    \label{fig:antenna-c-fits-data}
\end{figure}

Using the raw waveform data, we can attempt to constrain the amplitude of the isotropic galactic spectrum (see below for a discussion of this choice), if the noise level in any of the antennas is low enough. To that end, we remove the largest RFI or internal noise spikes using sigma-clipping with a threshold standard deviation of three. Any channel with PSD levels three standard deviations above or below the channel mean is thus removed.

We can then write the sigma-clipped PSD values from the waveform mode data as a matrix for each antenna, $S_\mathrm{ij}$, where the two subscripts $i,j$ index the packet in time (i.e. which of the 27 packets being analyzed) and by frequency channel, respectively.

To decrease the noise level, we average over all wave packets $N_t$ and average together different numbers of frequency channels $N_f$ within a final frequency bin $(\nu_k)$, where $k$ indexes the final bin:
\begin{equation}
    V_r^2(\nu_k) \equiv \frac{1}{N_f~N_t} \sum^{N_f}_j \sum^{N_t}_i S_\mathrm{ij},
\end{equation}
and where it is understood that all frequencies $j$ lie within the final bin $\nu_k$. Note that in our convention throughout this work, frequencies denoted by the Greek letter $\nu$ refer to those that have been averaged or are in MHz, while those denoted by the Latin letter $f$ refer to frequencies at native resolution. 

We take the variance in a single waveform from the ideal radiometer equation as 
\begin{equation}
    \sigma^2_\mathrm{base,ij} = \frac{S^2_{ij}}{\Delta f~\tau}
\end{equation}
where the dynamic range in the denominator is a function of the frequency resolution $\Delta f \approx 29296~$Hz and the integration time $\tau = 34.125~\mu s$. After averaging, by weighted propagation of errors, the variance in the PSD mean estimate becomes
\begin{equation}
\sigma^2_\mathrm{S} = \frac{\sum^{N_f}_j \sum^{N_t}_i \sigma^2_\mathrm{base,ij}} {N^2_t N^2_f} = \frac{\sum^{N_f}_j \sum^{N_t}_i S^2_{ij}} {\Delta f~\tau~N^2_t N^2_f},
\label{eqn-full-var}
\end{equation}
which, in the limit where the base variance $\sigma^2_\mathrm{base,ij}$ is a constant $\sigma^2_0$ that does not (significantly) vary between time packets, reduces to the familiar $1/\sqrt{N}$ form of the standard error of the mean. Given the limited number of wave packets available from ROLSES-1, we use Equation \ref{eqn-full-var} to estimate the variance in the waveform data.

Figure \ref{fig:antenna-c-fits-data} shows an example of data averaged in time and over seven adjacent frequency channels for Antenna C in black crosses, with corresponding error bars. After averaging, the error bars are $\sim 100~nV^2/Hz$, while the intrinsic scatter in the data is around $\sim 800 ~nV^2/Hz$. Note that the frequency range used here, $1.5-14~$MHz, was determined by analyzing the galactic spectrum as described in Section~\ref{sec:freq_range}. The blue and magenta curves and associated light blue and purple error bands in Figure \ref{fig:antenna-c-fits-data} show two different fits to the data in an attempt to constrain the isotropic galactic spectrum in the presence of electronic or receiver noise. These fits are described in detail in the next section.

\section{Galactic Spectrum and Antenna Parameter Fitting}
\label{sec:gal-spec-antenna}
\subsection{Antenna Parameters and Model}
The overall noise level in Antenna C is low enough to allow a model of the galaxy to be fit alongside key ROLSES-1 antenna parameters, such as the voltage divider or antenna gain factor $\Gamma$ and the effective length of the monopole antenna, $l_\mathrm{eff}$.

In general, the effective length of an antenna is related to the open-circuit voltage $V_\mathrm{oc}$ induced at the terminals by an impinging electric field $E(\boldsymbol{\hat{r}})$ as $l_\mathrm{eff}(\boldsymbol{\hat{r}}) = V_\mathrm{oc} / E(\boldsymbol{\hat{r}})$ \citep{balanis_antenna_2005}. Writing this as a function of a constant effective length that depends upon the (unknown) current distribution $l_\mathrm{eff}$ and directivity of the antenna $D(\boldsymbol{\hat{r}})$, or $l^2_\mathrm{eff}(\boldsymbol{\hat{r}}) = l^2_\mathrm{eff} D(\boldsymbol{\hat{r}})$, the squared voltage induced by the electric field at the receiver is 
\begin{equation}
    V^2_r = V^2_\mathrm{noise} + \frac{1}{2} \Gamma^2 l^2_\mathrm{eff} \int_{\Omega_s} E^2(\boldsymbol{\hat{r}}) D(\boldsymbol{\hat{r}}) d\Omega,
\end{equation}
where we have multiplied by $\Gamma^2$ to convert from open-circuit to received voltage, a factor of one half to account for averaging over all linear polarizations, and we have integrated over the solid angle of the source $\Omega_s$ whence the electric field comes \cite[see for instance][]{balanis_antenna_2005, zaslavsky_antenna_2011, condon_essential_2016, rolla_instrument_2024}. We have also added a constant noise offset term, $V^2_\mathrm{noise}$. In terms of brightness from a spatially constant source, $E^2 = B_{\nu}~Z_0$, where $Z_0$ is the impedance of free-space. Therefore, our equation becomes
\begin{equation}
    V^2_r = V^2_\mathrm{noise} + \frac{1}{2} \Gamma^2 l^2_\mathrm{eff} B_{\nu} Z_0 \int_{\Omega_s} D(\boldsymbol{\hat{r}}) d\Omega.
\end{equation}
If our source is isotropic over the entire sky ($\Omega_s = 4\pi$) and our antenna described by a dipole directivity pattern $D(\boldsymbol{\hat{r}}) \propto \sin^2{\theta}$, then our equation reduces to the form found in \cite{zaslavsky_antenna_2011} for measuring the isotropic galactic background using the free-space STEREO spacecraft antennas. 

However, for ROLSES-1 the monopole antennas are not dipoles poised in free-space, but monopole antennas situated above (or near) the lunar surface. The electrical conductivity of the lunar surface is relatively low, $\epsilon_r \approx1 - 2$ \citep{olhoeft_dielectric_1975,heiken_book-review_1991,li_lunar_2022}, and given the uncertainty in the exact orientation of Antenna C relative to the lunar surface, we approximate it as a dipole antenna positioned horizontally above the non-conducting lunar regolith and receiving brightness from half the sky, or $\Omega_s = 2\pi$. Our received voltage is then
\begin{equation}
    V^2_r = V^2_\mathrm{noise} + \frac{2 \pi}{3} Z_0 \Gamma^2 l^2_\mathrm{eff} B_{\nu}.
    \label{eqn:rec-psd}
\end{equation}
It is likely that the amount of brightness received is in reality less than that of a dipole detecting half the sky, as the horizon, large boulders, and the spacecraft itself may block radiation, as well as reduce the directivity of the antennas in general. Without more accurate knowledge of the conditions of the spacecraft and its lunar environment, however, and given that we have only five minutes of waveforms to work with, we assume that these other potentially parasitic effects are absorbed into the unknown parameters, $V^2_\mathrm{noise}$ and $\Gamma l_\mathrm{eff}$. For the latter, since the gain factor $\Gamma$ and effective length $l_\mathrm{eff}$ appear degenerately, we combine them into a single parameter, $\alpha = \Gamma^2 l^2_\mathrm{eff}$. 

\subsection{Galaxy Spectrum Model}
Many models of the low-frequency sky exist, including those constructed from principle component analysis or fitting of published all-sky maps \citep{de_oliveira-costa_model_2008,zheng_improved_2017,dowell_lwa1_2017,cong_ultralong-wavelength_2021}, or those parametrizing the low-frequency isotropic component of the galactic spectrum (i.e. the galactic poles) as a combination of synchrotron emission from galactic and extragalactic sources attenuated by free-free absorption \citep{cane_spectra_1979,novaco_nonthermal_1978, manning_galactic_2001}. Ideally the latter isotropic galactic spectrum component can be used as an absolute calibration source for the antenna parameters, as is done in \cite{dulk_calibration_2001, zarka_jupiters_2004, bale_electric_2008, zaslavsky_antenna_2011, page_l_2022, bassett_constraining_2023}. The assumption of isotropy is especially valid in a frequency range around $3~$MHz where free-free absorption by electrons in the plane of the galaxy cause the plane and poles to have the same brightness, as seen by e.g. low-gain antennas \citep{dulk_calibration_2001}. Above $3~$MHz, the plane is brighter than the poles and must be included as a separate component from the isotropic spectrum contributing to the total galactic brightness detected. However, this deviation from isotropy caused by the galactic plane is small at low frequencies, and even at $10~$MHz the difference between the two components is $\sim 5 \times 10^{-18}~V^2/Hz$ \citep[see Figure 1 for instance in][]{dulk_calibration_2001} for an electrically short monopole antenna. This deviation from isotropy is an order of magnitude smaller than the intrinsic scatter in the LS2 data points shown in Figure \ref{fig:antenna-c-fits-data} for antenna C, approximately $\sim 2 \times 10^{-17}~V^2/Hz$. Therefore, we assume that our model for the galactic spectrum is well approximated by its isotropic component in our frequency ranges of interest and for the ROLSES-1 waveform data.

Once the antenna parameters are known or constrained, then models of the antenna's beam or directivity pattern can be coupled with models of the galactic spectrum in order to reconstruct spatial maps of the latter \cite[again, see for example][]{page_l_2022, bassett_constraining_2023}. As only five minutes of raw waveform data are available from ROLSES-1, we shall attempt only to constrain the antenna parameters and amplitude of the isotropic galactic background component to within reasonable limits. The model of the galactic spectrum that we shall adopt is that of \cite{novaco_nonthermal_1978}, which was constructed from measurements made by the RAE-2 spacecraft:
\begin{equation}
    B_{\nu} = B_0 \nu^{-0.76} \exp{(3.28 \nu^{-0.64})} \equiv B_0 G(\nu)
    \label{eqn:gal-spec-nb}
\end{equation}
where $\nu$ are frequencies in MHz, and the amplitude parameter $B_0 = 1.38 \times 10^{-19}~W/m^2~Hz~sr$. We define $G(\nu)$ here to contain the shape information of the spectrum for ease of notation in later sections. The amplitude $B_0$ is the parameter that we shall allow to vary in our final fit to the ROLSES-1 Antenna C waveform data. 

\subsection{Frequency Range of the Model}
\label{sec:freq_range}

\begin{figure}
    \centering
    \includegraphics[width=0.48\textwidth]{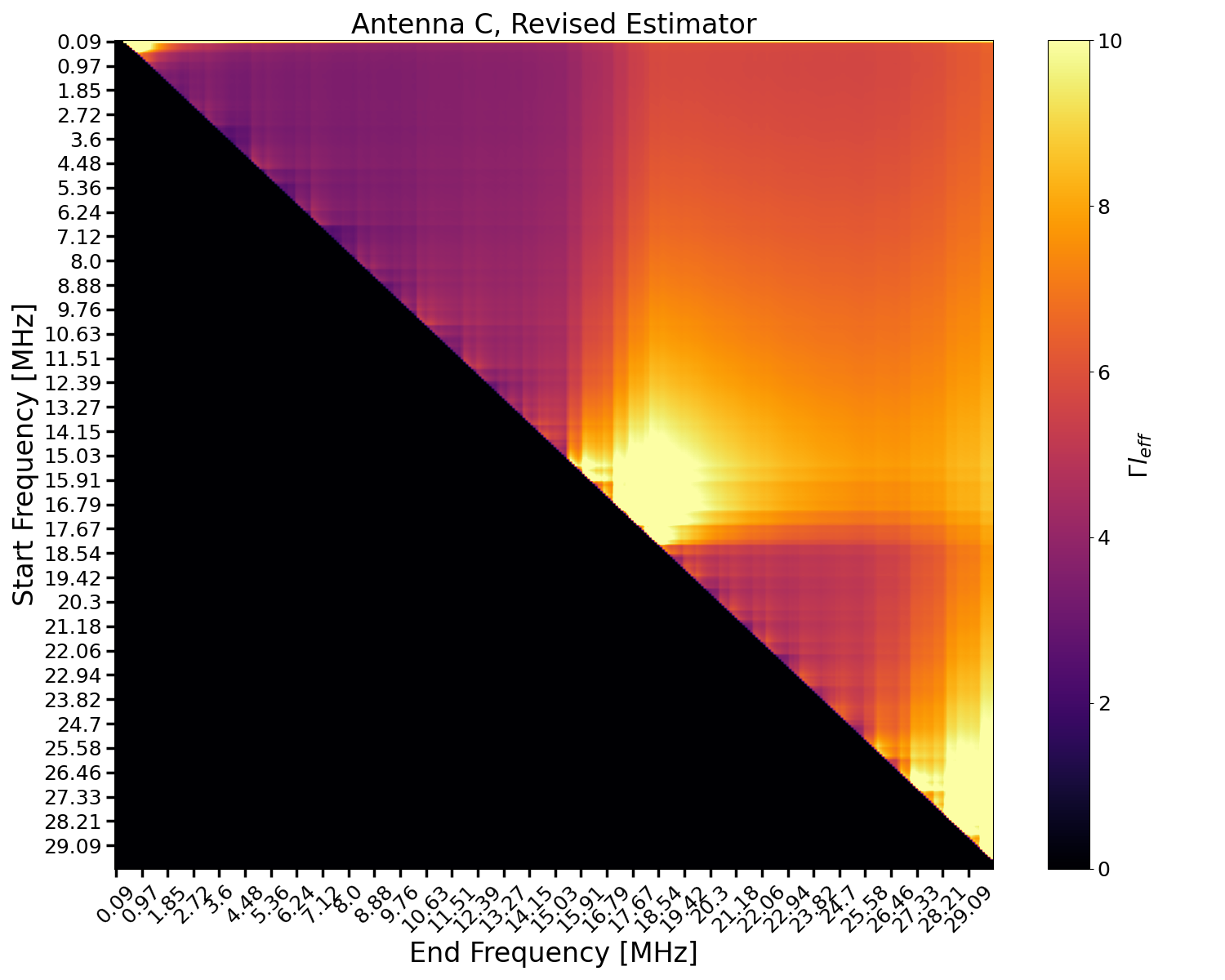}
    \caption{Estimate of the antenna parameters combination $\Gamma l_\mathrm{eff}$ over various frequency ranges using the signal amplitude estimator from Equation \ref{eqn:alpha-hat}. The y-axis represents the starting frequency over which the amplitude estimator is calculated, while the x-axis shows the end frequency used. This estimator assumes that the signal template of interest is the galactic spectrum of \cite{novaco_nonthermal_1978}, and that this signal is present in each spectrum. The estimator uses the LS2 raw waveform data from Antenna C to estimate $\Gamma l_\mathrm{eff}$. In this figure, the data show that this antenna parameters combination is relatively constant and small within the 1.5 - 15 MHz range. In general, the darker areas (purple and black) correspond to lower noise levels.}
    \label{fig:freq-range-estimator}
\end{figure}

In various cases, it is possible to use empirical arguments to determine a more precise frequency range over which the model in Equation \ref{eqn:rec-psd} applies, given the implicit assumption that both the noise response $V^2_\mathrm{noise}$ and the combined antenna parameter $\Gamma l_\mathrm{eff}$ are frequency independent. For instance, in \cite{zaslavsky_antenna_2011} the frequency range is determined by arguing that above $4~$MHz the short dipole approximation breaks down for the STEREO antennas, while below several hundred kHz quantum thermal noise dominates and must be included as a term. For the ROLSES-1 monopole antennas, $L=2.5~m \ll \lambda$ for even the highest frequency observed by both the DSP and waveform data at 30 MHz, or $\lambda \approx 10~$m. Quasi-thermal noise tends to dominate in the kHz frequency range, but its magnitude especially is dependent upon the local plasma environment \citep{meyer-vernet_quasi-thermal_2017}. For the Moon, an estimate of the electron density ($N_e \approx 300 ~cm^{-3}$) and electron temperature ($T_e \approx 2.2~$eV) from \cite{farrell_lunar_2013} gives a PSD noise contribution below 750 kHz of roughly $\sim 10^{-15}~V^2/Hz$ for $l_\mathrm{eff} \approx L/4$ for a monopole antenna. Therefore, for our galactic spectrum analysis, we focus on frequencies above 1 MHz to avoid this source of noise. 

In order to determine the optimal frequency range for our model, and in particular where the antenna parameter combination is constant, we employ an analysis technique from 21-cm cosmology, specifically from \cite{switzer_erasing_2014}. In the latter work, the authors derive analytical estimates for an unknown amplitude parameter $\alpha$ which is multiplying a known spectral template $\boldsymbol{x}$, where the template is a column vector. The analysis assumes that the template form is known a \textit{a priori}, but that its amplitude lies far below all systematics and sources of noise. In \cite{switzer_erasing_2014}, the signal template is assumed to be the global 21-cm signal; in our analysis we shall assume it is the isotropic galactic spectrum of \cite{novaco_nonthermal_1978}, and that the amplitude parameter (multiplying the spectrum) we wish to estimate is the following squared combination of unknown antenna parameters: $\alpha = \Gamma^2 l^2_\mathrm{eff}$.

The estimate of the amplitude $\hat{\alpha}$ is constructed by first assuming that the signal in question consists of two components, one parallel to the noise and systematics, and one orthogonal to them: $\boldsymbol{x} = \boldsymbol{x}_{\parallel S} + \boldsymbol{x}_{\perp S}$. We seek to project out the component perpendicular to the systematics. To find that component, we first calculate the estimate of the spectral covariance matrix of the data:
\begin{equation}
    \boldsymbol{\hat{\Sigma}} = \frac{1}{N_t - 1} \sum^{N_t}_i \left( \boldsymbol{S}_i - \boldsymbol{\bar{S}} \right) \left( \boldsymbol{S}_i - \boldsymbol{\bar{S}} \right)^T
    \label{eqn:cov-matrix}
\end{equation}
where $\boldsymbol{S}_i$ denotes, as before, each spectrum labeled by a time index $i$ and $\boldsymbol{\bar{S}}$ is the time-averaged spectrum. The largest eigenmodes $\boldsymbol{f}_k$ of this covariance matrix $\boldsymbol{\hat{\Sigma}}$ can then be used to project out the signal component perpendicular to the systematics and give us an estimate of the amplitude of the desired signal template via
\begin{equation}
    \hat{\alpha} = \frac{\boldsymbol{x}^T(1-\boldsymbol{F}\boldsymbol{F}^T)\boldsymbol{\bar{S}}}{\boldsymbol{x}^T(1-\boldsymbol{F}\boldsymbol{F}^T)\boldsymbol{x}},
    \label{eqn:alpha-hat}
\end{equation}
where the matrix $\boldsymbol{F}$ contains the largest systematic eigenmodes $\boldsymbol{f}_k$ as columns \cite[note this is the same as Equation 29 in][]{switzer_erasing_2014}. We can use these modes because the isotropic galaxy appears in every spectrum $\boldsymbol{S}_i$, and so the subtraction of the time-averaged spectrum removes the contribution of the galaxy from the covariance matrix. This is desirable, as otherwise our eigenmodes from this covariance matrix would, after subtraction, remove power from the galactic spectrum. Mathematically, the numerator in this equation finds the components of the data parallel to the systematic eigenmodes and subtracts them out before projecting the rest into the signal template space, while the denominator applies the appropriate normalization given the eigenmodes and signal template. Note that this equation implies that components of the systematics (e.g., from combinations of electronic receiver noise or thermal sky noise) that are parallel to the signal will be leftover after the subtraction, and thus the estimate $\hat{\alpha}$ may contain residual systematic power.

Although Equations \ref{eqn:cov-matrix} and \ref{eqn:alpha-hat} are written for the entire frequency spectrum, we can truncate the spectrum vectors $\boldsymbol{S}_i$ to whatever range we desire before calculating the covariance and amplitude estimator. If we find that the estimate of the signal amplitude $\hat{\alpha} = \Gamma^2 l^2_\mathrm{eff}$ changes when calculated over a larger frequency range, this would imply that there is a greater contribution to the systematic component parallel to our signal template (either from new or increased sources of electronic noise, RFI, etc.) present in that range, or that our assumption that $\Gamma l_\mathrm{eff}$ is constant over that particular frequency range is invalid. We therefore calculate Equation \ref{eqn:alpha-hat} over a variety of frequency ranges using the data $\boldsymbol{S}_i$ from Antenna C and a signal template for the galactic spectrum in PSD:
\begin{equation}
    \boldsymbol{x} = \frac{2 \pi Z_0}{3} \boldsymbol{B}_{\nu},
\end{equation}
where $\boldsymbol{B}_{\nu}$ is the isotropic galaxy spectrum of \cite{novaco_nonthermal_1978}, as shown in Equation \ref{eqn:gal-spec-nb}. The factor multiplying the signal template is necessary to convert the galaxy spectrum from brightness to PSD, after multiplication by the amplitude estimate $\hat{\alpha}$.

Figure \ref{fig:freq-range-estimator} shows a range of values for the square-root of the amplitude estimator, $\sqrt{\hat{\alpha}} = \Gamma l_\mathrm{eff}$, obtained when calculating Equation \ref{eqn:alpha-hat} over differing frequency intervals. The y-axis denotes the starting frequency while the x-axis denotes the end frequency assumed in the calculation, which specifically fixes the indices $j$ in $S_{ij}$. The amplitude appears relatively small and constant over the frequency range $\sim 1 - 15~$MHz before dramatically increasing, probably due to the presence of increased electronic noise or RFI in the data. The purple-shaped wedge in the $1-15~$MHz band, however, gives us confidence that the model is relatively constant in this range, regardless of what are chosen as starting and stopping frequencies (unless a very small range of frequencies is chosen, in which case noise fluctuations dominate).

The absolute values of our estimator here are larger ($\sim 2-3$ times) than we would expect based upon theoretical and experimental considerations for the individual values of $\Gamma$ and $l_\mathrm{eff}$. The former can in general never exceed unity, while the latter should be $\leq L = 2.5$~m, which is the physical length of the monopole antennas for ROLSES-1. The value we have estimated for $\hat{\alpha}$ thus probably includes a small component of power parallel to the systematics. In addition, this estimate is extremely sensitive to the absolute value of the data $\boldsymbol{S}_i$, as can be seen from Equation \ref{eqn:alpha-hat}. Small calibration errors will therefore bias the amplitude estimate. In the absence of better methods to calibrate the raw waveform data, however, we employ here these amplitude estimates.

From this analysis, we conservatively choose to apply our model of the galactic spectrum over the range $1.5 - 14~$MHz. Moreover, we have checked that our conclusions do not change qualitatively if we use smaller frequency ranges. Given that frequency range, we can calculate a final value for our estimator using a jack-knife analysis. For this, we recompute the amplitude estimator in Equation \ref{eqn:alpha-hat} after removing individual data points, giving us pseudo-values for the amplitude estimate. After iterating through the entire set of data points, we then calculate the average of all the pseudo-values to find the final jack-knife mean and variance, $\hat{\alpha}_\mathrm{JK}$ and $\hat{\sigma}^2_\mathrm{JK}$, respectively. Figure \ref{fig:antenna-c-fits-data} shows in blue our estimate for the galactic spectrum mean with a 1-$\sigma$ error band (light blue) in power spectral density space, for which our jack-knife values of the amplitude estimator are multiplied by the assumed galaxy spectrum signal template, $\hat{\alpha}_\mathrm{JK} \boldsymbol{x}$. We note the bias in the mean and the large size of the errors, and take this result as an upper limit for our estimate of the antenna parameters. In the next section, we describe our Bayesian posterior sampling analysis to estimate the parameters of the full model, including the antenna parameters, noise, and galactic spectrum amplitude. 

\subsection{Bayesian Posterior Estimation of the Full Model}
Nested sampling algorithms such as \texttt{PolyChord} \citep{handley_polychord_2015} allow us to numerically explore complex, high-dimensional parameter spaces using prior distribution knowledge of the input parameters and a likelihood distribution. Using the reduced data from Antenna C, $V^2_r(\nu_k)$, which we denote in vector form as $\boldsymbol{V}^2_r$, we can write the log likelihood, or the conditional probability of the data given the parameters, in the usual way as
\begin{equation}
    \boldsymbol{\mathcal{L}}(\boldsymbol{V}^2_r|\boldsymbol{\theta}) \propto -\frac{1}{2} \left[\boldsymbol{V}^2_r - \boldsymbol{M}_g(\boldsymbol{\theta})\right]^T \boldsymbol{C}^{-1} \left[\boldsymbol{V}^2_r - \boldsymbol{M}_g(\boldsymbol{\theta})\right],
\end{equation}
where $\boldsymbol{C}^{-1}$ is the noise covariance, and our model for the data follows Equation \ref{eqn:rec-psd} as 
\begin{equation}
    \boldsymbol{M}_g(\boldsymbol{\theta}) = V^2_\mathrm{noise} \boldsymbol{1} + \frac{2 \pi}{3} Z_0 \Gamma^2 l^2_\mathrm{eff} B_0 \boldsymbol{G}
\end{equation}
and $\boldsymbol{G}$ again contains all of the spectral information of the galactic spectrum of \cite{novaco_nonthermal_1978}, and $\boldsymbol{1}$ is simply a column vector of ones with a length equal to the number of frequencies. The parameter vector of this model $\boldsymbol{\theta}$ is $[V^2_\mathrm{noise}, \Gamma l_\mathrm{eff}, B_0]$. We set uniform priors on all of the parameters, as shown in Table \ref{tab:pars-priors}, and sample the $\Gamma l_\mathrm{eff}$ parameter in $\log_\mathrm{10}$-space. Although the antenna parameters combination $\Gamma l_\mathrm{eff}$ and the galactic amplitude $B_0$ appear multiplicatively within the same term--and hence constitute a degeneracy--we regulate this degeneracy by enforcing strict prior ranges on the parameters. 

Finally, we note that the $1\sigma$ error bars on the individual data points in Figure \ref{fig:antenna-c-fits-data} are small compared to the intrinsic scatter of the data, even after averaging in time and frequency. The source of this scatter is likely due to large noise modes present after averaging, an unknown systematic modulation on top of the (assumed) constant electronic noise and galactic spectrum, or thermal fluctuations due to the small amount of data available to us. Therefore, to model the effects of this scatter, we add an additional intrinsic scatter noise term to our total error budget and vary it alongside the other parameters. The total variance of our model is then
\begin{equation}
    \sigma^2_T = \sigma^2_\mathrm{S} + \sigma^2_\mathrm{int}
\end{equation}
where $\sigma_\mathrm{S}$ is the statistical variance from Equation \ref{eqn-full-var} including the effects of averaging, and $\sigma_\mathrm{int}$ is the noise parameter we allow to vary. Our full covariance matrix for the likelihood is then a diagonal matrix with $\boldsymbol{C} = \textbf{diag}(\sigma^2_T)$. After the nested sampling fit, we then marginalize over the noise parameter $\sigma_\mathrm{int}$ to obtain our final posterior mean estimates for the other parameters. With our likelihood, priors, and parameters in hand, we can then sample from our posterior distribution $ P(\boldsymbol{\theta}|\boldsymbol{V}^2_r)$ using Bayes' theorem and \texttt{PolyChord}:
\begin{equation}
    P(\boldsymbol{\theta}|\boldsymbol{V}^2_r) = \frac{\boldsymbol{\mathcal{L}}(\boldsymbol{V}^2_r|\boldsymbol{\theta}) \boldsymbol{\pi}(\boldsymbol{\theta})}{Z}
\end{equation}
where $\boldsymbol{\pi}$ are the prior distributions on the parameters, and $Z$ is the Bayesian evidence.

\begin{table}[]
    \centering
    \begin{tabular}{c|c}
        \hline
        Parameter & Prior Range [Units] \\
        \hline 
        \hline
        $V^2_\mathrm{noise}$ & Uniform(0,10) $[10^{-16} V^2/Hz]$ \\
        $\Gamma l_\mathrm{eff}$ & $Log_\mathrm{10}$ Uniform(-3, 0.3) [m] \\
        $B_0$ & Uniform(-220, -160) [dB] \\
        $\sigma_\mathrm{int}$ & Uniform(0,10) $[10^{-16} V^2/Hz]$ \\
    \end{tabular}
    \caption{Prior ranges used for each of the parameters explored with the Bayesian nested sampling algorithm. }
    \label{tab:pars-priors}
\end{table}

\section{Results}
\label{sec:results}

\begin{figure*}
    \centering
    \includegraphics[width=0.98\linewidth]{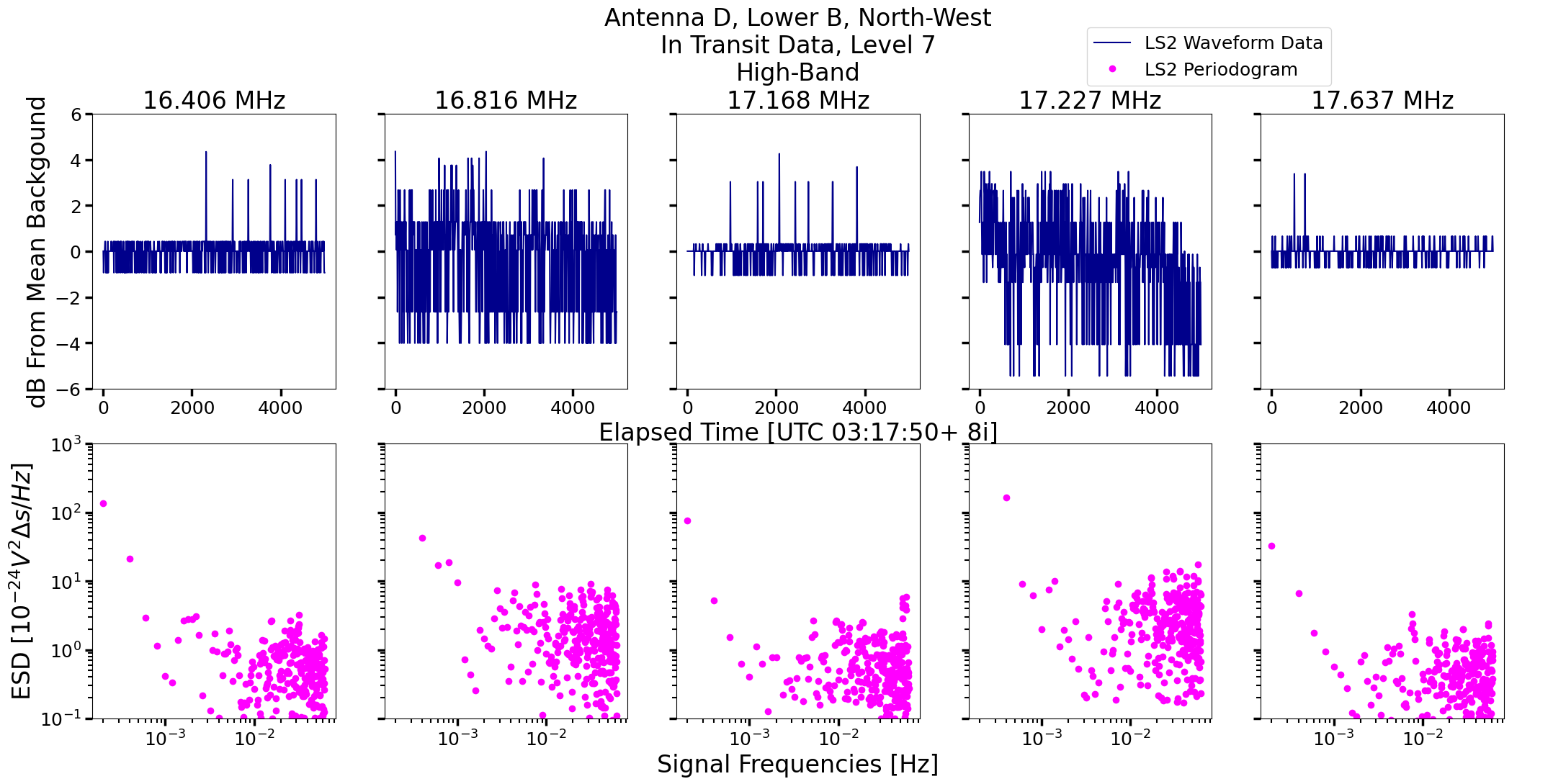}
    \caption{Spectra and periodograms for several frequency channels with candidate shortwave transmissions (technosignatures) for antenna D during the In-Transit observation day from the DSP. The top row shows the spectra in decibels relative to its own average value, while the bottom row shows the periodogram for each corresponding channel in the top row. The x-axis for the fluctuations in the top panel are indexed by $i$ and are each eight seconds apart, while the x-axis for the periodograms are signal frequencies in Hertz. Note the level of fluctuations in the spectra as well as the dominant periods in the periodograms: the latter exhibit decreasing power modulation with increasing frequency modulation, as expected for shortwave transmissions breaking through the Earth's ionosphere.}
    \label{fig:dsp-db-and-periodogram}
\end{figure*}

\subsection{DSP Data and Technosignatures}

We speculate that the strong PSD lines detected by the DSP and waveform data to be the result of shortwave transmissions generated by stations and satellites on Earth breaking through the ionosphere--or ``technosignatures.'' The plasma cutoff frequency for the night-side ionosphere is typically around $\sim 5$~MHz, but can be as low as $1-2$~MHz, and most signals detected by ROLSES-1 (and correspondingly by LWA at the same local time) are far above this cutoff value. These shortwave breakthroughs appear to ``twinkle'' with timescales on the order of minutes. If these signals are shortwave breakthroughs, then we can assume the transmitters to be quasi-uniform sources on the timescales observed by ROLSES-1 (an hour or less), with the power varying only as a function of the cosine of the angle between the antenna and Earth (which also does not significantly change on these timescales). The scintillations we observe on minute timescales would instead be due to modulation by the ionosphere as waves of electron density enhancements pass over the various transmitters \citep{beniguel_ionospheric_2019, fallows_lofar_2020}. Data taken by LWA on the same day at the same time show similar shortwave transmission lines (see Appendix \ref{app-lwa}) most densely in the $14-23~$MHz region with a drop-off of signals below $\sim 11~$MHz. The higher frequency lines are also seen by Radio Jove (see Gopalswamy et al. 2025 in prep). 

Figure \ref{fig:dsp-db-and-periodogram} shows the spectra (top row) and periodograms (bottom row) for several of the channels with the strongest PSDs from the In-Transit data. The spectra in blue are shown in decibels relative to the average PSD value for that channel, where different channels are separated by column. Time on the x-axis for the spectra is shown as seconds past the observation start time, indexed by $i$ and taken every eight seconds. The bottom row shows the corresponding periodogram for each spectrum; we calculate the periodogram by multiplying each spectrum in each channel by a 4th order Blackman-Harris window function before taking the DFT. The squared absolute value of the DFT is shown in magenta with the x-axis specifying the frequencies of the signal. As can be seen from the periodograms, most of the power in each spectrum corresponds to periods on the order of minutes, with periodogram power decreasing as a function of increasing modulation frequency. Furthermore, the spectra in the top row have fluctuations on the order of $5-10$~dB. All of these features---the minute timescale for fluctuations, the $5-10$~dB range of spectra fluctuations, and the decrease in scintillation modulation with increasing modulation frequency---are characteristic of shortwave transmissions breaking through the ionosphere \citep{kaiser_windwaves_1996, beniguel_ionospheric_2019}. The scintillation modulation power is strongest in the 17.227 MHz observation, with enhanced modulation values at and below 1 mHz as compared to the other observations (see bottom row of Figure \ref{fig:dsp-db-and-periodogram}). The low frequency modulation of the signal is even seen as ``M-shaped'' maxima and minima values in the measurements shown in the top panel for 17.227 MHz. The channel at 16.816 MHz also shows similar modulation. We note that similar patterns are observed for the waveform data when we examine spectra in channels individually and with their corresponding periodograms, although the latter are less informative as we have only five minutes of data and are thus sensitive only to timescales up to five minutes.

The observation of the ionospheric scintillations in the shortwave transmissions suggests a possible future space weather application: Obtaining a terrestrial map of scintillation activity from a Moon-based radio observatory. Specifically, if the short wave transmitter locations were known, one could then create a scintillation modulation index for the ionospheric region above the transmitter. From a platform on the Moon, the index can be mapped across the Moon-facing surface of the Earth to provide hemispherical coverage of changes in the index.

\subsection{Waveform Data and the Galactic Spectrum} 
\label{sec:modeling-and-fits}

\begin{figure}
    \centering
    \includegraphics[width=0.5\textwidth]{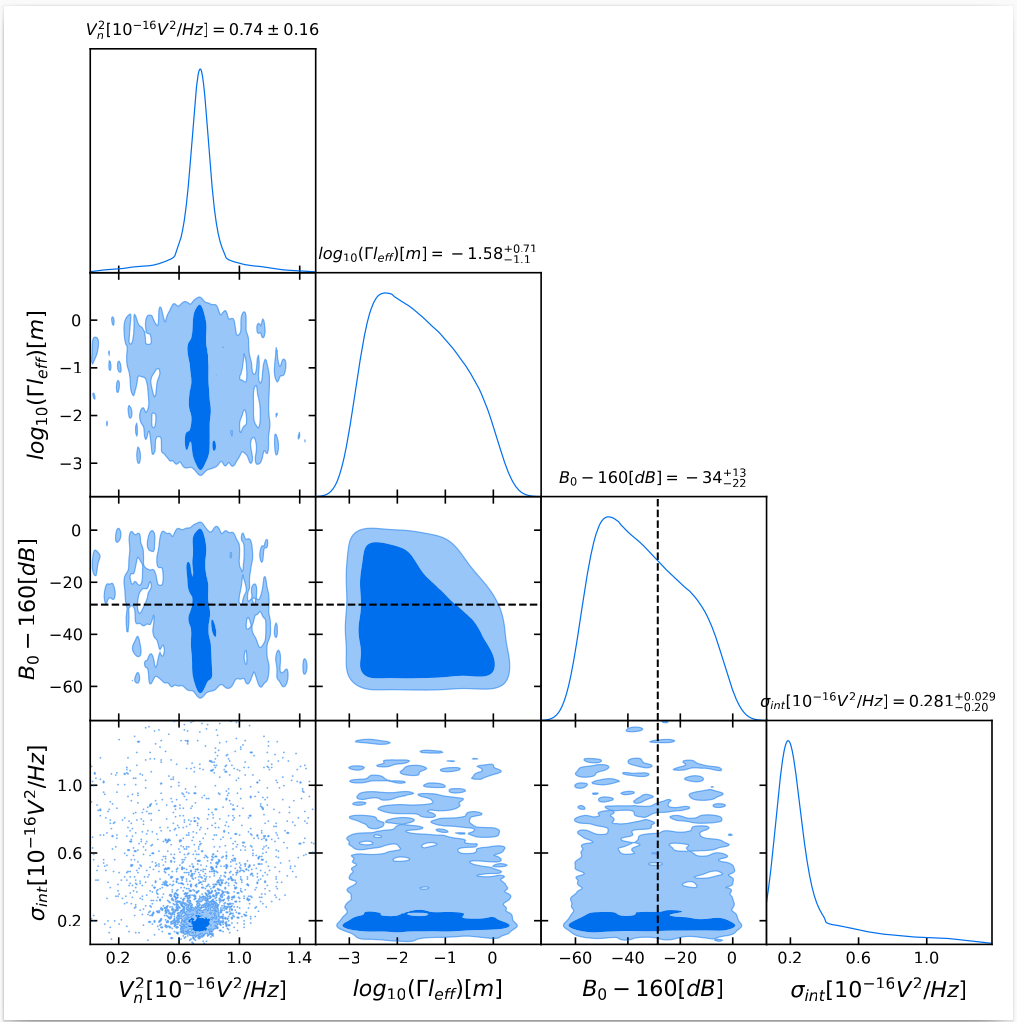}
    \caption{Triangle plot showing the extracted covariances ($68\%$ and $95\%$ confidence intervals in dark and light blue, respectively) on the parameters of our Antenna C receiver response model, Equation \ref{eqn:rec-psd}. The noise parameters describing the mean electronic noise and intrinsic scatter about the mean are $V^2_n$ and $\sigma_\mathrm{int}$, respectively. $\Gamma l_\mathrm{eff}$ represents the antenna parameters combination converting brightness of sources to PSD and is sampled in $\log_{10}$-space, while $B_0$ is the amplitude of the isotropic galactic spectrum. The black dashed line shows the fiducial galactic amplitude value of \cite{novaco_nonthermal_1978}.}
    \label{fig:gal-triangle}
\end{figure}


Our main result from our efforts to detect the isotropic galactic spectrum from the five minutes of raw waveform data of Antenna C is shown in Figure \ref{fig:gal-triangle}, a triangle plot depicting the posterior covariances for the various parameters. The contours show the two dimensional covariances between parameters, while the one dimensional plots show the marginalized densities for each parameter. Above each one dimensional plot is shown the marginalized parameter value with $68\%$ confidence intervals. In addition, Figure \ref{fig:antenna-c-fits-data} shows in magenta (line and error band) our best-fit full model $M_g$ from using the mean posterior parameter values and their $1\sigma$ confidence intervals.

The noise parameters $V^2_\mathrm{noise}$ and $\sigma_\mathrm{int}$ in Figure \ref{fig:gal-triangle} are sharply peaked around values which correspond to the mean electronic noise (and therefore offset of the data, regardless of calibration), and the intrinsic scatter around this mean. Moreover, the antenna parameters combination is surprisingly independent of the galactic amplitude $B_0$, and exhibits a mean value of $\mu_{\Gamma l_\mathrm{eff}} = 10^{-1.58} \sim 0.026$, although the confidence intervals are large. Importantly, this value is well within theoretical expectations, as we discuss below. Similarly, the galactic amplitude parameter exhibits a reasonable mean value of $\mu_{B_0} = -194~dB = 3.98 \times 10^{-20}~W/m^2~Hz~sr$ (where the value in dB is relative to $1~V^2/Hz$), with the original amplitude parameter of \cite{novaco_nonthermal_1978} indicated by dashed black lines in Figure~\ref{fig:gal-triangle}. Again, the marginalized confidence intervals around the mean value are large, yet even in the face of strong parameter degeneracies, large noise levels, and only a modest amount of data, our methods have allowed a value for the amplitude of the galactic spectrum compatible with previous measurements to be extracted from the ROLSES-1 data.

In addition, we ran another nested sampling fit with the same prior ranges for the noise parameters $V^2_\mathrm{noise}$ and $\sigma_\mathrm{int}$, but instead did not allow the antenna parameter combination or galactic spectrum amplitude to vary, constituting a \textit{noise-only} fit. As \texttt{PolyChord} produces the Bayesian evidence, we can use this statistic to determine model preference, either for a fit with only noise parameters or for one with noise and antenna plus galactic spectrum parameters. Our noise-only fit for antenna C gives a Bayesian evidence of $\log_{10}(Z_\mathrm{noise}) = 1.494$, while the fit including the galactic spectrum yields $\log_{10}(Z_\mathrm{gal}) = 1.554$, meaning there is a slight preference for the fit containing the galactic spectrum, with odds of $\sim 1.0618 : 1$.  If we restrict our model to a smaller frequency range of $1.5 - 6~$MHz, this comparison produces a bit larger odds, $\sim 1.14 : 1$. 

As a second test, we also calculate the Kolmogorov-Smirnoff (K-S) p-value \citep{kolmogorov_sulla_1933,smirnov_table_1948} for a particular fit by computing the K-S statistic from a comparison of the cumulative distribution function for a unit normal distribution to the best-fit model residuals normalized by the total noise (including the intrinsic scatter noise). A model which leaves only Gaussian residuals at the noise level of the fit is consistent with the ``true'' model \citep[see Section 4 of][for an more exhaustive elaboration of this idea]{hibbard_fitting_2023}, and will therefore be preferred over models which leave residuals above the noise level. Calculating the K-S p-value for our galactic spectrum fit, we find $p_\mathrm{gal} = 0.753$. Given the null hypothesis that our model leaves only Gaussian residuals at the noise level, and with a standard significance threshold of $p=0.05$, we can conclude that our model that includes the galactic spectrum passes the null hypothesis. We note that the noise-only fit K-S value also passes the null hypothesis, albeit with only a slightly lower value than the galactic spectrum fit ($p_\mathrm{noise}=0.752$). 

Finally, as a last consistency check, we calculate fits in the same manner as antenna C for the other waveform data from antennas A and B to determine if they are consistent with noise-only fits or with data containing the isotropic galactic spectrum. In particular, given the high noise level in antenna A relative to B and C, we expect preference to be given to a noise-only fit. The estimators for Antenna A indicate a preferred frequency range around $3.2-12~$MHz, with amplitude $\alpha_A \geq 20$, while antenna B is preferred only in the $10-22~$MHz range with amplitude estimate $\alpha_B \sim 8$. Evidently the low frequencies of antenna B are corrupted with noise or other features parallel to the signal template, although the amplitude estimate is smaller than antenna A given that the latter is a low gain antenna (see Table \ref{tab:data}). 

After sigma-clipping and averaging seven adjacent frequency bins together (to facilitate comparison with the antenna C results), we find for antenna A that the Bayesian evidence actually favors a model that includes the galactic spectrum versus a noise-only fit ($-242$ versus $-1086$, respectively), but that neither of the models pass the K-S test significance threshold ($3\times10^{-4}$ versus $2\times10^{-215}$ respectively), and therefore we reject the null hypothesis that only Gaussian residuals at the noise level remain after the fit. For antenna B, we find that the Bayesian evidence and K-S p-values are nearly identical for the galactic spectrum and noise-only fits, with values of $-117$ and $0.92$, respectively. There is thus no statistical preference, even slight, for the galactic spectrum fit over a noise-only fit, although both fits are consistent with Gaussian residuals at the noise level of the data. 

\subsubsection{Antenna Parameters}

Attempting to infer independent values for the antenna parameters is impossible in this formalism, as they are modeled as a single parameter due to their degeneracy. Laboratory testing or independent, additional measurements would be required to break this degeneracy. In lieu of such circumstances, we are left only with a reasonable estimate based upon similar antenna designs. Monopole antennas are often assumed to have an effective length of roughly $l_\mathrm{eff} \sim L/4$ \citep{balanis_antenna_2005}, which is $\sim 0.625$ for the $L=2.5~$m antennas of ROLSES-1. Using this value and the mean posterior estimate for $\Gamma l_\mathrm{eff}$, we can then infer that the antenna gain factor is $\Gamma \sim 0.042$. 

In these radio frequency ranges, the resistive parts of the antennas are negligible, making them purely a function of the antenna capacitance $C_a$ and the stray capacitance $C_s$ \citep{balanis_antenna_2005, zaslavsky_antenna_2011, rolla_instrument_2024} through the following formula:
\begin{equation}
    \Gamma = \frac{C_a}{C_a + C_s}.
\end{equation}

The antenna capacitance for thin linear wire antennas can be treated with the analytical formula
\begin{equation}
    C_a = \frac{2 \epsilon_0 \pi}{k} \frac{\tan{(kL)}}{\log{(L/a)}-1}
\end{equation}
where $\epsilon_0$ is the vacuum permittivity, $k$ is the wavevector, and $a$ is the radius of the antenna. The factor of two accounts for the fact we are modeling the monopole as a single antenna above an infinite ground plane. 

Using the average antenna radius for $a$, we find an antenna capacitance of roughly $C_a \sim 15~$pF for the ROLSES-1 antenna C and which is very weakly dependent on frequency. This estimate for the antenna capacitance then gives a stray capacitance value due to the environment and spacecraft of $C_s \sim 340~$pF, although we again emphasize that this is only the mean value, and the spread around this will be quite large given the uncertainty in the combination $\Gamma l_\mathrm{eff}$. Such a large value however is not perhaps surprising, given the potential parasitic effects of the lunar environment and spacecraft. A more detailed analysis, however, would require vastly more data, a demand that greatly exceeds the capabilities of the now deceased lander.

Lastly, we note that this capacitance analysis only applies to the antennas on the lunar surface, where the lunar environment likely increases the local stray capacitance, and cannot therefore be directly applied to the In-Transit data. The latter data, taken in the magnetosheath, will exhibit a different local capacitance, and a more precise calibration of the DSP data would require taking these local capacitance losses into account. 

\section{Conclusions} 
\label{sec:conclusions}

In defiance of a litany of obstacles, the Intuitive Machines lunar lander \textit{Odysseus} landed on the South pole of the Moon near crater Malapert A. On board the lander rode NASA's first radio telescope to the Moon, called \textit{Radiowave Observations on the Lunar Surface of the photo-Electron Sheath}-1 (ROLSES-1) consisting of four stacer monopole antennas observing in the ranges $0.06-30$~MHz and $0.06 - 1.8~$MHz, high and low band, respectively. Data were collected from one rogue antenna of ROLSES-1 en route to the Moon in the high band, and after landing the three remaining antennas were deployed and collected data as well. In total, ROLSES-1 observed on three separate days, one In-Transit and two on the lunar surface, totaling around two hours of data from four antennas. Most of the data came from the digital signal processor (DSP) in the form of spectra, but five minutes of raw waveforms were also telemetered back to Earth, giving two different modes of data to process and analyze. Unfortunately, after processing, only high band data remained. 

The DSP data were cleaned from all known sources of noise, referencing both ground test data in addition to that from stowed antennas after launch, in order to determine channels contaminated by electronic noise from the lander and ROLSES-1 power source. Sources of noise inherent to the DSP itself were also examined and removed and finally the data were calibrated using laboratory test data and converted to power spectral density, giving approximately $\sim 100 ~$ minutes of data for the In-Transit day and the second day upon the lunar surface. The five minutes of raw waveforms captured on the lunar surface during the second day of observations were converted to voltages using the nominal gain requirements for the antennas and on-board electronics before being transformed to spectra using standard signal processing. After applying these cleaning steps and calibration techniques, we find good agreement between the PSD levels generated from the DSP versus those calculated from the raw waveforms, for all antennas which collected both modes of data on the lunar surface.

The channels which remain after the extensive cleaning process contain bright PSD lines especially in the $14 - 24~$MHz range which we speculate are shortwave transmissions breaking through the Earth's ionosphere, or terrestrial technosignatures, as viewed from ROLSES-1. Such shortwave transmissions are quasi-uniform sources on the timescales of the ROLSES-1 observations, so all modulation in their magnitude is likely due to the ionosphere in the form of electron density enhancements passing over the transmitters on Earth. This causes a scintillation effect with a periodicity of the order of several minutes and corresponding periodograms which have decreasing power modulation with increasing frequency. Compared to experiments utilizing other spacecraft observing in similar frequency bands \cite[see e.g.,][]{kaiser_windwaves_1996}, shortwave transmissions also exhibit $5-10$~dB-level fluctuations in amplitude, again due to the effects of the ionosphere. The PSD lines from ROLSES-1 exhibit all three of these effects in both the DSP and waveform modes of data. Finally, data taken at approximately the same time with the Long Wavelength Array in Owens Valley and with the Radio Jove experiment in Mexico show similar shortwave transmission features across the observing band. Therefore, we find it likely that ROLSES-1 measured anthropogenic radiowaves or shortwaves breaking through the Earth's ionosphere and escaping to the Moon. Such data may also be of interest for research aimed at modeling potential technosignatures as they would appear from distant exoplanets with Earth-like technology \citep{sagan_search_1993,johnson_simultaneous_2023}.

Lastly, after extensive sigma-clipping and averaging, we used the five minutes of raw waveform data from the least noisy ROLSES-1 monopole--Antenna C--to perform a first-order calibration of combined antenna parameters, including the antenna gain factor and effective length, using the low-frequency isotropic galactic background spectrum as a calibration source. After building a model of the antenna receiver response that includes noise parameters and the galactic background spectrum of \cite{novaco_nonthermal_1978}, and utilizing techniques from 21-cm cosmology to determine the frequency range of the model, we used Bayesian analysis and the nested sampling algorithm \texttt{PolyChord} to explore the parameter space of our model. The resultant covariances constrained both the antenna parameter combination and the amplitude of the isotropic galactic spectrum to within reasonable limits, given theoretical expectations of the former and a fiducial value of the latter taken from the well-known galaxy spectrum of \cite{novaco_nonthermal_1978}. A subsequent run of the nested sampling algorithm which included only noise parameters was slightly disfavored over the aforementioned fit that included the isotropic galactic spectrum, according to both the Bayesian evidence and the K-S p-value of the noise-normalized residuals.

ROLSES-2, the follow-up mission, will launch in approximately two years and will be poised to make unprecedented measurements of the low-frequency radio sky, lunar environment, auroral kilometric radiation from the Earth, solar bursts, and potentially Jovian emission. Equipped with an on-board calibration source, direction-finding capability, and with all four Stokes parameters, ROLSES-2 will perhaps be able to stand tall upon the leaning shoulders of ROLSES-1. Another lunar radio telescope called LuSEE-Night is set to launch in early 2026 and is poised to place the first constraints on the 21-cm signal of the cosmic Dark Ages. Both of these experiments will face calibration and data reduction problems similar to ROLSES-1, and thus much of this work has been devoted to developing tools and techniques for analyzing lunar radio telescope data. 

\textbf{ACKNOWLEDGMENTS}
We first would like to thank many of the engineers on the ROLSES-1 team for invaluable knowledge about the engineering and data collection of ROLSES-1, including Richard Katz, Igor Kleyner, Richard Mills, and Jack Quire. We would like to thank Bang Nhan for useful discussions (and simulations) involving the ROLSES-1 antennas and analysis of the waveform data, and Stuart Bale for enlightening discussions concerning the noise levels present in the ROLSES-1 data and the capacitive coupling of ROLSES-1 to the lunar environment. We would also like to thank An\v{z}e Slozar for helpful discussions concerning the quantized noise level and DSP algorithm. In addition, we would like to thank Nivedita Mahesh, Greg Hellbourg, and Gregg Hallinan for providing the LWA data taken on 02/21/24. We acknowledge support by NASA grant 80NSSC23K0013. The data analysis for ROLSES-1 was also funded by NASA grant 80NSSC22K1264 from the Exploration Science Strategy Integration Office (ESSIO). We wish to thank Joel Kearns for his support of ROLSES-1 and his leadership of the CLPS program. J.D.T. was supported for this work by the TESS Guest Investigator Program G06165 and by NASA through the NASA Hubble Fellowship grant $\#$HST-HF2-51495.001-A awarded by the Space Telescope Science Institute, which is operated by the Association of Universities for Research in Astronomy, Incorporated, under NASA contract NAS5-26555.

\appendix

\section{PSD Normalizations from DFTs}
\label{app-psd}

Let the DFT of an N point time-domain function $x[n]$ multiplied by a windowing function $w[n]$ be given by 
\begin{equation}
    \tilde{X}[f] = \frac{1}{N} \sum^N_{n=0} x[n]w[n]e^{-2 \pi i n f /N}
\end{equation}
where $f$ is the corresponding point in the frequency domain. The power spectral density is always defined as the normalized (i.e. divided by $N$) power spectrum divided by the frequency resolution $\Delta f$ (sometimes called the bandwidth), where $\Delta f = 1 / \Delta t$, and $\Delta t$ is the spacing between points in the time domain. For the DFT, the following relations hold:
\begin{equation}
    T = N \Delta t, \Delta f = \frac{f_s}{N}, 
\end{equation}
where $f_s$ is the sampling frequency and $T$ is the total time of the signal with $N$ points. The power spectral density is then
\begin{equation}
    S_w[f] \equiv \frac{2}{\Delta f} \frac{1}{W_\mathrm{ss}} |\tilde{X}[f]|^2 = \frac{2N}{f_s} \frac{1}{W_\mathrm{ss}} |\tilde{X}[f]|^2
\end{equation}
\cite[see][]{paschmann_analysis_1998}. Here the factor of $2$ arises as we are only including positive frequencies in our definition of the DFT from $n=0$.

This is the normalization convention for the DFT used by IDL. For Matlab, Python, and Simulink, the DFT is defined without the $1/N$ factor \citep{paschmann_analysis_1998}. That is,
\begin{equation}
    \tilde{X}_\mathrm{python}[f] = \sum^N_{n=0} x[n]w[n]e^{-2 \pi i n f /N},
\end{equation}
so that the PSD now becomes
\begin{align}
    S_w[f] = \frac{2N}{f_s} \frac{1}{W_\mathrm{ss}} |\tilde{X}[f]|^2 = \frac{2N}{f_s} \frac{1}{W_\mathrm{ss}} \frac{1}{N^2} |\tilde{X}_\mathrm{python}[f]|^2 = \frac{2N}{f_s} \left( \frac{1}{N} \sum^N_{n=0} w^2[n]\right)^{-1} \frac{1}{N^2} |\tilde{X}_\mathrm{python}[f]|^2 \\ = \frac{2}{f_s} \left(\sum^N_{n=0} w^2[n]\right)^{-1} |\tilde{X}_\mathrm{python}[f]|^2 = \frac{2T}{N} \left(\sum^N_{n=0} w^2[n]\right)^{-1} |\tilde{X}_\mathrm{python}[f]|^2.
\end{align}

\section{Owens Valley Long Wavelength Array (LWA) Data}
\label{app-lwa}

\begin{figure*}
    \centering
    \includegraphics[width=0.48\textwidth]{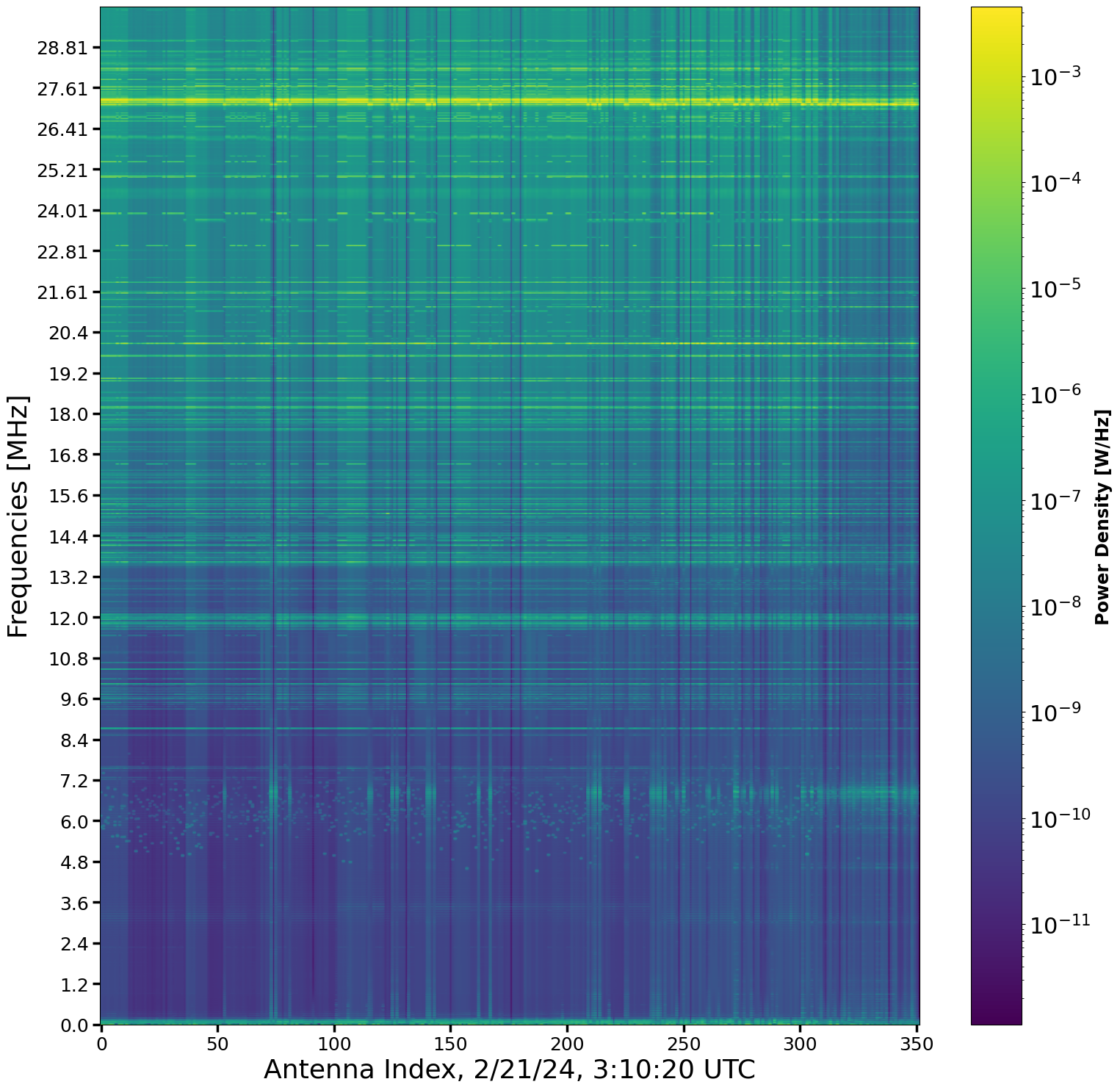}
    \caption{The figure shows the power density (W/Hz) values recorded by all 352 LWA antennas on 2/21/24 at 3:10:20 UTC, right before ROLSES-1 collected data during its cruise phase. Frequencies are shown on the y-axis in MHz, while each antenna is indexed by the x-axis. The data from LWA have been provided courtesy of Nivedita Mahesh, Greg Hellbourg, and Gregg Hallinan.}
    \label{fig:lwa-data}
\end{figure*}

Figure \ref{fig:lwa-data} shows the autocorrelation data taken by all 352 LWA antennas on day 2/21/24 at the single time 3:10:20~UTC, with the x-axis denoting each individual antenna and the y-axis denoting frequencies in MHz. The color scale indicates power density with units of $W/Hz$, and we remind the reader that this is not technically a waterfall plot, as it shows all autocorrelations seen by each antenna at one time. Such a snapshot is generated automatically by the observatory every three hours, and the transmission lines shown in the figure appear extremely variable, since the next snapshot at 6:10:20~UTC shows many different transmission lines from those shown here. Nonetheless, we note that most of the antennas detected bright lines in the $\sim16- 23~$MHz band, and that there appears to be a similar drop-off of lines and signal strength below $\sim11~$MHz as that seen by ROLSES-1. The LWA data therefore confirm the general trends seen by ROLSES-1 during its In-Transit observations.

Making a comparison between the two sets of data in order to e.g. calculate the power of the transmitters, however, is complicated by the fact that the paths and hence the attenuation factors between the transmitters and the two observatories are extremely different and unknown. The LWA is likely seeing a combination of power from a transmitter due to free-space-loss, refraction, reflection of the ionosphere, ducting, etc., while ROLSES-1 sees the power that breaks through the ionosphere. In addition, ROLSES-1 sees the globally-averaged view of shortwave transmitters.

Using the Friis equation, we can however calculate an approximate value for the power of the shortwave transmitters $P_t$ detected by ROLSES-1:
\begin{equation}
    P_{r}(\theta, \phi, \lambda) = P_t D_r(\theta_r,\phi_r) D_t(\theta_t,\phi_t) \left(\frac{\lambda}{4 \pi d} \right)^2,
    \label{eq:friis}
\end{equation}
where the power received by a ROLSES-1 antenna $P_r$ depends upon the power of the transmitter $P_t$, the directivities of the receiver and transmitter, $D_t$ and $D_r$ respectively, the wavelength $\lambda$ of the transmitted signals, and the distance $d$ between the receiving and transmitting antennas \citep{balanis_antenna_2005}. This equation assumes that the reflection efficiencies of both antennas are unity, and the polarization loss factor and efficiency are unity (i.e., the polarizations are matched between the receiving antenna and transmitted signal). The received power $P_r$ we take from the time-averaged In-Transit ROLSES-1 data, and examine specifically the three frequencies with bin centers which overlap the most with LWA, which are $\nu_i = [18.46, 21.15, 23.55]~$MHz. We assume in our analysis that the transmitter directivity $D_t$ is isotropic and therefore unity, a fair assumption for transmitters that are FM radio or broadcast stations. Note that we also assume that the power loss is dominated by the free-space-loss factor scaling as $1/d^2$, although there is probably a small amount of power loss due to the ionosphere as well (as we see evidence of scintillations, as stated above). 

For the In-Transit ROLSES-1 data, we assume that the antenna can be approximated as a short dipole in free space, and therefore the directivity is $D_r(\theta)\sim 1.5 \sin^2{\theta}$ \citep{balanis_antenna_2005}. Ignoring less bright sources, we can then integrate $P_{r}$ over the solid angle of the Earth as seen by ROLSES-1 from $\sim 3.89\times 10^8~$m away, which is approximately $1.83^{\circ}$ in circular diameter. Carrying out the integral, we find that the directivity (assuming the Earth is roughly perpendicular to the antenna) is $D_r \sim 1.5 \times 0.2$.

Solving Equation \ref{eq:friis} for $P_t$, we find for the three frequencies that
\begin{equation}
    P_t \approx [99.4, 25.9, 294.9]~ \text{kW}
\end{equation}
for the transmitting power of the shortwave transmitters detected by ROLSES-1 in cis-lunar space. For the brightest lines at $\nu_i = [16.82, 17.23, 18.4]$, we find 
\begin{equation}
    P_t \approx [9.1,13.4,16.7]~ \text{MW}.
\end{equation}
Again we note that our estimates for the brightest lines are off by a small factor of $\sim 5$ or so, probably due to the uncertainty in the zero-level of the calibration sweep data for the DSP. See Section \ref{sec:data} for a discussion of the calibration.

\bibliography{references}{}
\bibliographystyle{aasjournal}

\end{document}